\documentclass[twocolumn, tighten]{aastex62}
\usepackage{natbib,amssymb,amsmath,graphicx,here}
\usepackage{hyperref}
\begin{document}

\title{The LEECH Exoplanet Imaging Survey: Limits on Planet Occurrence Rates under Conservative Assumptions}

\author[0000-0003-0454-3718]{Jordan M. Stone}
\altaffiliation{Hubble Fellow} 
\affiliation{Steward Observatory,
University of Arizona,
933 N. Cherry Ave,
Tucson, AZ 85721-0065 USA} 

\author{Andrew J. Skemer}
\affiliation{Department of Astronomy and Astrophysics, 
University of California, Santa Cruz, 
1156 High St, 
Santa Cruz, CA 95064, USA}

\author{Philip M. Hinz}
\affiliation{Steward Observatory,
University of Arizona,
933 N. Cherry Ave,
Tucson, AZ 85721-0065 USA} 

\author{Mariangela Bonavita}
\affiliation{SUPA, Institute for Astronomy, University of Edinburgh, Royal
Observatory, Edinburgh EH9 3HJ, UK}

\author[0000-0001-5253-1338]{Kaitlin M. Kratter}
\affiliation{Steward Observatory,
University of Arizona,
933 N. Cherry Ave,
Tucson, AZ 85721-0065 USA} 

\author{Anne-Lise Maire}
\affiliation{Max Planck Institute for Astronomy, K{\"o}nigstuhl 17, D-69117
Heidelberg, Germany}

\author{Denis Defrere}
\affiliation{Space sciences, Technologies \& Astrophysics Research (STAR)
Institute, University of Li\`ege, Li\`ege, Belgium}

\author{Vanessa P. Bailey}
\affiliation{Jet Propulsion Laboratory, California Institute of Technology,
Pasdena, CA, 91109}

\author[0000-0003-3819-0076]{Eckhart Spalding}
\affiliation{Steward Observatory,
University of Arizona,
933 N. Cherry Ave,
Tucson, AZ 85721-0065 USA} 

\author{Jarron M. Leisenring}
\affiliation{Steward Observatory,
University of Arizona,
933 N. Cherry Ave,
Tucson, AZ 85721-0065 USA} 

\author{S. Desidera}
\affiliation{INAF Osservatorio Astornomico di Padova, vicolo dell’ Osservatorio
5, 35122 Padova, Italy}

\author{M. Bonnefoy}
\affiliation{Univ. Grenoble Alpes, CNRS, IPAG, F-38000 Grenoble, France}

\author{Beth Biller}
\affiliation{SUPA, Institute for Astronomy, University of Edinburgh, Royal
Observatory, Edinburgh EH9 3HJ, UK}
\affiliation{Centre for Exoplanet Science, University of Edinburgh, Edinburgh
EH9 3HJ, UK}
  
\author{Charles E. Woodward}
\affiliation{Minnesota Institute of Astrophysics, 
University of Minnesota,
116 Church Street, SE,
Minneapolis, MN 55455, USA}

\author{Th. Henning}
\affiliation{Max Planck Institute for Astronomy, K{\"o}nigstuhl 17, D-69117
Heidelberg, Germany}

\author{Michael F. Skrutskie}
\affiliation{Department of Astronomy, University of Virginia, Charlottesville,
VA 22904, USA}

\author[0000-0002-1031-4199]{J. A. Eisner}
\affiliation{Steward Observatory,
University of Arizona,
933 N. Cherry Ave,
Tucson, AZ 85721-0065 USA} 

\author{Justin R. Crepp}
\affiliation{Department of Physics, University of Notre Dam, 225 Niuwland
Science Hall, Notre Dame, IN, 46556, USA}

\author{Jennifer Patience} 
\affiliation{School of Earth and Space Exploration, Arizona State University,
PO Box 871404, Tempe, AZ 85287, USA}

\author[0000-0001-9754-2233]{Gerd Weigelt} 
\affiliation{Max-Planck-Institut f{\"u}r Radioastronomie,
Auf dem H{\"u}gel 69, 53121 Bonn, Germany}

\author{Robert J. De Rosa}
\affiliation{Astronomy Department, University of California, Berkeley, CA
94720, USA}

\author{Joshua Schlieder}
\affiliation{Exoplanets and Stellar Astrophysics Laboratory, Code 667, NASA
Goddard Space Flight Center, 8800 Greenbelt Rd., Greenbelt, MD 20771, USA}
\affiliation{Max Planck Institute for Astronomy, K{\"o}nigstuhl 17, D-69117
Heidelberg, Germany}

\author{Wolfgang Brandner}
\affiliation{Max Planck Institute for Astronomy, K{\"o}nigstuhl 17, D-69117
Heidelberg, Germany}

\author{D{\'a}niel Apai}
\affiliation{Steward Observatory, University of Arizona, 933 N. Cherry Ave,
Tucson, AZ 85721-0065 USA}
\affiliation{Lunar and Planetary Laboratory, The University of Arizona, 1629 E.
Univ. Blvd., Tucson, AZ 85721 USA}
\affiliation{Max Planck Institute for Astronomy, K{\"o}nigstuhl 17, D-69117
Heidelberg, Germany}

\author{Kate Su}
\affiliation{Steward Observatory, University of Arizona, 933 N. Cherry Ave,
Tucson, AZ 85721-0065 USA}

\author{Steve Ertel}
\affiliation{Steward Observatory,
University of Arizona,
933 N. Cherry Ave,
Tucson, AZ 85721-0065 USA} 

\author{Kimberly Ward-Duong}
\affiliation{Five College Astronomy Department, Amherst College, Amherst, MA
01002, USA}

\author[0000-0002-1384-0063]{Katie M. Morzinski}
\affiliation{Steward Observatory,
University of Arizona,
933 N. Cherry Ave,
Tucson, AZ 85721-0065 USA} 

\author{Dieter Schertl}
\affiliation{Max-Planck-Institut f{\"u}r Radioastronomie,
Auf dem H{\"u}gel 69, 53121 Bonn, Germany}

\author{Karl-Heinz Hofmann}
\affiliation{Max-Planck-Institut f{\"u}r Radioastronomie,
Auf dem H{\"u}gel 69, 53121 Bonn, Germany}

\author{Laird M. Close}
\affiliation{Steward Observatory,
University of Arizona,
933 N. Cherry Ave,
Tucson, AZ 85721-0065 USA} 

\author[0000-0002-1440-3666]{Stefan S. Brems}
\affiliation{Zentrum f{\"u}r Astronomie der Universit{\"a}t Heidelberg,
Landessternwarte, K{\"o}nigstuhl 12, 69117 Heidelberg, Germany}

\author{Jonathan J. Fortney}
\affiliation{Department of Astronomy and Astrophysics, 
University of California, Santa Cruz, 
1156 High St, 
Santa Cruz, CA 95064, USA}

\author{Apurva Oza}
\affiliation{Department of Astronomy, University of Virginia, Charlottesville,
VA 22904, USA}
\affiliation{Physikalisches Institut, Universit{\"a}t Bern, Gesellschaftsstrasse 6,
CH-3012 Bern, Switzerland}

\author{Esther Buenzli}
\affiliation{Institute for Particle Physics and Astrophysics, ETH Zurich,
Wolfgang-Pauli-Strasse 27, 8093 Zurich, Switzerland}

\author{Brandon Bass}
\affiliation{Steward Observatory,
University of Arizona,
933 N. Cherry Ave,
Tucson, AZ 85721-0065 USA} 

\begin{abstract} We present the results of the largest $L^{\prime}$
($3.8~\mu$m) direct imaging survey for exoplanets to date, the Large Binocular
Telescope Interferometer (LBTI) Exozodi Exoplanet Common Hunt (LEECH). We
observed 98 stars with spectral types from B to M. Cool planets emit a larger
share of their flux in $L^{\prime}$ compared to shorter wavelengths, affording
LEECH an advantage in detecting low-mass, old, and cold-start giant planets.
We emphasize proximity over youth in our target selection, probing physical
separations smaller than other direct imaging surveys. For FGK stars, LEECH
outperforms many previous studies, placing tighter constraints on the hot-start
planet occurrence frequency interior to $\sim20$~au.   For less luminous,
cold-start planets, LEECH provides the best constraints on giant-planet
frequency interior to $\sim20$~au around FGK stars. Direct imaging survey
results depend sensitively on both the choice of evolutionary model (e.g., hot
or cold-start) and assumptions (explicit or implicit) about the shape of the
underlying planet distribution, in particular its radial extent.  Artificially
low limits on the planet occurrence frequency can be derived when the shape of
the planet distribution is assumed to extend to very large separations, well
beyond typical protoplanetary dust-disk radii ($\lesssim50$~au), and when
hot-start models are used exclusively. We place a conservative upper limit on
the planet occurrence frequency using cold-start models and planetary
population distributions that do not extend beyond typical protoplanetary
dust-disk radii. We find that $\lesssim90\%$ of FGK systems can host a 7 to 10
$M_{\mathrm{Jup}}$ planet from 5 to 50~au. This limit leaves open the
possibility that planets in this range are common.  \end{abstract}


\section{Introduction} 

Understanding the formation and evolution of planetary systems requires
a detailed census of the galactic planet population.  Surveys sensitive to
planets orbiting at a wide range of semi-major axes about stars with a variety
of spectral types and ages are required to measure the planet mass and
semi-major axis distributions as a function of host mass, age, metallicity, and
environment.

The Doppler spectroscopy and transit photometry techniques have excelled at
discovering mature planets on short-period orbits about their field-aged (few
Gyr) host stars \citep[e.g.,][]{Cumming2008, Batalha2014}.  Microlensing
surveys have helped constrain the planet population around typically older
low-mass stars \citep[e.g.,][]{Gould2010}.  The direct imaging technique is
sensitive to wide-orbit planets typically around younger stars
\citep[e.g.,][]{Bowler2016}.  In this paper, we report the results of the Large
Binocular Telescope Interferometer (LBTI) Exozodi Exoplanet Common Hunt
\citep[LEECH;][]{Skemer2014b} direct imaging survey of intermediate-aged stars.  

The direct imaging technique of planet detection entails high spatial
resolution adaptive optics (AO) assisted observations and sophisticated
post-processing techniques to separate starlight from planet light
\citep[e.g.,][]{Marois2006, Soummer2012}. Since direct imaging involves
collecting photons directly from the planetary photosphere, the atmospheres of
any newly discovered planets can be studied in detail --- potentially leading to
constraints on composition and the formation process \citep{Konopacky2013,
Barman2015, Skemer2016, Samland2017}.

Past direct imaging surveys \citep[recently reviewed in][]{Bowler2016} probed
orbits with semi-major axes $\gtrsim10$~au (with peak sensitivity
$\gtrsim100$~au) for nearby young stars (with typical distances $\sim50$~pc),
filling in some of the parameter space not covered by radial-velocity or
transit surveys. Probing these outer regions is important because this is the
regime where protoplanetary disks are thought to have enhanced surface density
of solids due to the freeze-out of volatiles, adding to the raw material
available to build up protoplanetary cores \citep[e.g.,][]{Hayashi1981}.
Therefore, direct imaging provides access to the semi-major axes where giant
planets could form in-situ \citep[e.g.,][]{Ormel2010, Lambrechts2012}. 

The results of previous direct imaging surveys are conveniently tabulated in
\citet{Chauvin2015} and \citet{Galicher2016}. \citet{Bowler2016} provides an
overview and summary of previous imaging surveys and synthesizes their results
in an analysis that combines data from many studies. The main result is that
gas-giant planets on orbits $\gtrsim100$~au are rare. As we shall show,
interior to $\sim100$~au constraints on the planet occurrence frequency are
less well established and depend on the assumed shape of the planet population
distribution, as well as the initial entropy and luminosity evolution of giant
planets. In particular, a cold-start population that does not extend beyond
typical protoplanetary dust-disk radii remains poorly constrained.

Younger and more massive planets are easier to detect with direct imaging,
because planets --- lacking an internal heat source --- cool and fade as they
radiate their gravitational potential energy. As a result, most direct imaging
surveys emphasize youth in their target selection to maximize their sensitivity
to low-mass gas-giant planets. Selecting for youth is an absolute necessity
for surveys conducted at wavelengths $<2.5~\mu$m where planet-to-star contrast
is a steep function of the planetary effective temperature
\citep[e.g.,][]{Liu2010, Chauvin2015, Galicher2016, Tamura2016}. However, the
planet-to-star contrast is alleviated at longer wavelengths near the peak of
a planet's spectral energy distribution.  Furthermore, previous discoveries
have shown that near-infrared ($\lambda\lesssim2.5~\mu\mathrm{m}$) emission is
suppressed and thermal-infrared ($3-5~\mu\mathrm{m}$) emission is enhanced for
directly imaged planets compared to more-massive brown dwarfs at the same
effective temperature, a consequence of low surface gravity
\citep[e.g.,][]{Chauvin2005, Skemer2011, Barman2011,Marley2012,Filippazzo2015}.
We conducted the LEECH survey in the $L^{\prime}$ band at 3.8~$\mu$m to take
advantage of an increased sensitivity to cooler planets, including lower-mass
planets, older planets, and planets that are born cold because they accrete
their envelopes through a radiatively efficient shock
\citep[e.g.,][]{Marley2007}. We therefore emphasize proximity over youth in our
target selection to probe the smallest orbital distances possible given the
constraints of diffraction-limited observing at $3.8~\mu$m with an 8.25~m
aperture.

\begin{figure*}
\includegraphics[width=\textwidth]{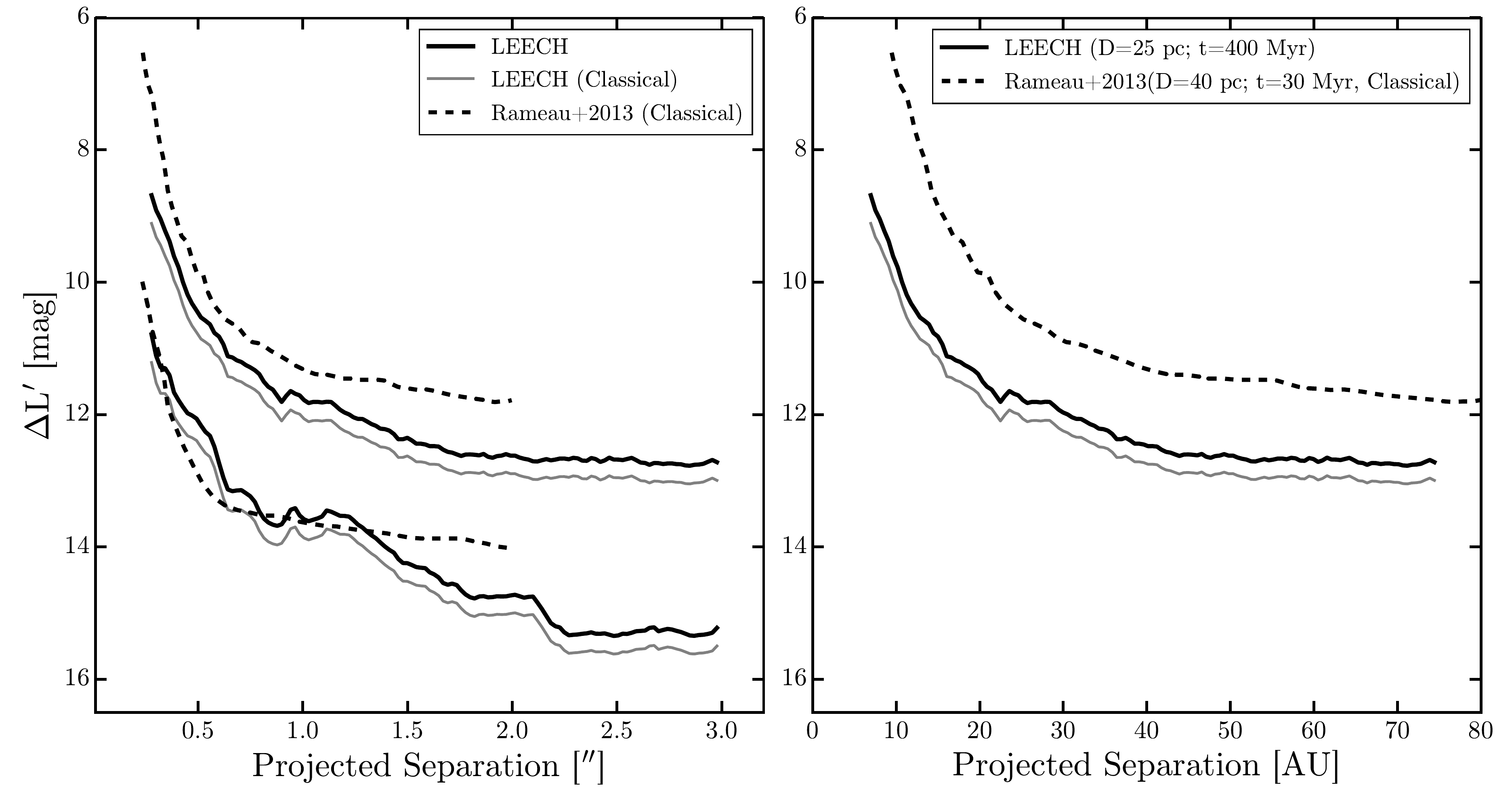}
\caption{The median and best LEECH contrast curves and comparison curves from
the $L^{\prime}$ direct imaging survey by \citet{Rameau2013}.  Left: contrast
as a function of projected angular separation. We show two sets of contrast
curves for the LEECH survey. With black solid lines we show the contrast curves
generated using a more modern approach that more carefully quantifies our
confidence limits. With gray lines we show our LEECH contrast curves using
a classical 5$\sigma$ definition. The gray curves can be directly compared to
the black dashed curves representing the \citet{Rameau2013} survey performance.
Right: Contrast versus projected separation in astronomical units for a median contrast at each
survey's median distance. A typical target for the LEECH survey closer by,
so we probe deeper at similar physical separations.\label{contrastcurves}}
\end{figure*}

\citet{Kasper2007} were the first to perform an $L^{\prime}$ direct imaging
survey, observing 22 stars with the Very Large Telescope (VLT) before the
advent of modern high-contrast post-processing algorithms. Additional
$L^{\prime}$ surveys are summarized in Table \ref{surveysSummary}. Our survey
includes many of the same targets as the \citet{Heinze2010A} survey, and we
probe deeper and closer in to the host star to search for lower-mass planets on
shorter period orbits. In Figure \ref{contrastcurves} we compare the median and
best LEECH contrast curves from LBTI/LMIRCam to the median and best contrast
curves from the \citet{Rameau2013} survey, which used NACO at the VLT and
delivered the deepest contrasts among all the NACO surveys listed in Table
\ref{surveysSummary}.  The LEECH survey stands out among earlier direct imaging
surveys at $L^{\prime}$, having the largest target list and best median
contrast.

\begin{deluxetable*}{lccccc}
\tabletypesize{\footnotesize}
\tablecolumns{6}
\tablewidth{0pt}
\tablecaption{Summary of $L^{\prime}$ Surveys\label{surveysSummary}}
\tablehead{
    \colhead{Reference}& 
    \colhead{Instrument}&
    \colhead{Number of}&
    \colhead{Spectral}&
    \colhead{Median Distance}&
    \colhead{Median Age}\\
    \colhead{} &
    \colhead{} &
    \colhead{Targets}&
    \colhead{Types}&
    \colhead{(pc)}&
    \colhead{(Myr)}
    }
\startdata
\citet{Kasper2007}  & NACO & 22 & FGKM   & 37   & 10-30               \\
\citet{Heinze2010A} & Clio & 54 & FGKM   & 11.2 & $\sim200$           \\
\citet{Delorme2012} & NACO & 16 & M      & 25.4 & 12\tablenotemark{a} \\
\citet{Rameau2013}  & NACO & 59 & BAFGKM & 40   & 30                  \\
\citet{Meshkat2015} & NACO & 13 & AF     & 48   & 40                  \\
\citet{Lannier2016} & NACO & 58 & M      & 38   & 21                  \\
LEECH               & LBTI & 98 & BAFGKM & 25.5 & 400                 \\
\enddata 
\tablenotetext{a}{\citet{Delorme2012} assigned a uniform age of 12 Myr to members
of the $\beta$~Pic moving group, and objects from this group dominate their
target list, resulting in a 12 Myr median age. \citet{Lannier2016} assigned an
age of 21 Myr to these same targets.}
\end{deluxetable*}

Our $L^{\prime}$ sensitivity is facilitated by the unique architecture of the
LBT observatory and the LBTI instrument. The observatory provides twin
deformable secondary mirrors each with 672 actuators that routinely deliver
images with $\sim90\%$ Strehl at 3.8 $\mu$m \citep{Bailey2014}. By using the
secondary mirror as the adaptive element, the LBTAO system minimizes the number
of warm optics in the light path to provide low-background high-throughput
thermal and mid-infrared images.  LBTI was designed for thermal and
mid-infrared science and includes a cryogenic beam combiner for feeding the
light from each side of the LBT into the instrument and onto the detector
\citep{Hinz2016}. Observations for the LEECH survey were performed in direct
imaging mode with LBTI and are not interferometric.

We describe our target selection in Section \ref{TargetSelection}, where we
also derive ages for a subset of A- and B- type stars on our target list.  Our
observing strategy and data processing pipeline are described in Sections
\ref{Observations} and \ref{DataProcessing}, respectively. We present the
results of our survey in Section \ref{Results}. In Section \ref{Analysis}, we
perform a statistical analysis of the results. We provide a detailed discussion
about how different assumptions can significantly alter inferred limits on
wide-orbit planet occurrence rates in Section \ref{discussionSection}. We then
outline the conditions for a conservative estimate.

\section{Target Selection}\label{TargetSelection}

The LEECH survey was designed with two facets: (1) a statistical survey to help
better inform our understanding of the wide-orbit giant-planet distribution
and (2) characterization of known low-mass systems and directly imaged planets.
Objects specifically added to the LEECH target list for characterization
include HIP~72567 (HD~130948), HIP 114189 \citep[HR~8799, LEECH
reference:][]{Maire2015}, HIP~64792 \citep[GJ~504, LEECH
reference:][]{Skemer2016}, and HIP~116805 ($\kappa$~And).

Although characterization targets meet the selection criteria (described below)
for our statistical sample target lists, we leave them out of our statistical
analysis. These targets were specifically prioritized for observation under
good observing conditions and when adequate time could be committed to reach
the necessary sensitivity.  Because prioritization enhances our ability to
detect planets around stars with known companions, including them in our
statistical analysis would bias our results.  The HIP 21547 (51 Eri) system
provides a good example of our reasoning. This system was on our original
target list and observed by LEECH before the discovery of a directly imaged
planet \citep{Macintosh2015}. As discussed below, we obtained only marginal
datasets on this star taken through thin clouds. We did not detect the planet.
Since the planet around 51 Eri was not known before we observed the system with
LEECH, we do include this object in our statistical analysis.
\citet{Schlieder2016} presented a characterization of the NO~UMa system using
LEECH data; however, this system was not specifically targeted for
characterization and is also included in our statistical sample.

\subsection{Master Target List for Statistical Survey}
We compiled a master target list comprising four sublists for use during the
LEECH survey.  Each sublist carried a slightly different emphasis, though the
guiding principles for each were relative proximity and age $\lesssim1$~Gyr.

Our first sublist emphasizes proximity and F/G/K spectral type. Targets for
this FGK sublist were drawn from \citet{Heinze2010B} and \citet{Mamajek2008}.
In total we observed 17 stars from the Heinze/Mamajek sublist. Our second
sublist is composed of stars in the Ursa Major moving group selected from
\citet{King2003}. This sublist provides a set of targets, with spectral types
ranging from A to M, that all have the same well-constrained age
\citep[$414\pm23$~Myr;][]{Jones2015}.  We observed 31 stars from the UMa
sublist. Our third sublist includes A- and F-type stars that show evidence of a debris
disk, drawn from \citet{Gaspar2013}. In total, we observed 17 stars from this
Dusty-A/F sublist.  Finally, our fourth sublist includes B- and A-type field
stars with estimated ages $\lesssim1$~Gyr (see Section \ref{fieldBAsubsec}).
We observed 33 stars from this Field-B/A sublist.

During observing nights, we selected targets from our master list based on  (1)
the total amount of parallactic angle change accessible in a three hour block 
and (2) the position of the target with respect to the wind velocity, which
restricts azimuth angles accessible for good adaptive-optics performance.
Stars were also prioritized by our best guess for the probability of hosting
a wide-orbit gas-giant planet based on planet frequency correlations with
host-star properties deduced from radial-velocity surveys \citep[e.g. mass and
metallicity;][]{Crepp2011}, although observing conditions at the telescope
typically drove nightly target selection.

Table \ref{TargetSummary} provides details on all targets observed during the
course of the LEECH survey. Stellar masses were derived by fitting to PARSEC
isochrones \citep{Marigo2017} using the target age and photometry, except for
the Field-B/A sublist for which mass and age were fit simultaneously (see
Section \ref{fieldBAsubsec}). Magnitudes in the $L^{\prime}$-band were derived
using the $K-L^{\prime}$ or $V-L^{\prime}$ color spectral-type relations
of \citet{Bessell1988}.

\begin{figure*} 
\includegraphics[width=\textwidth]{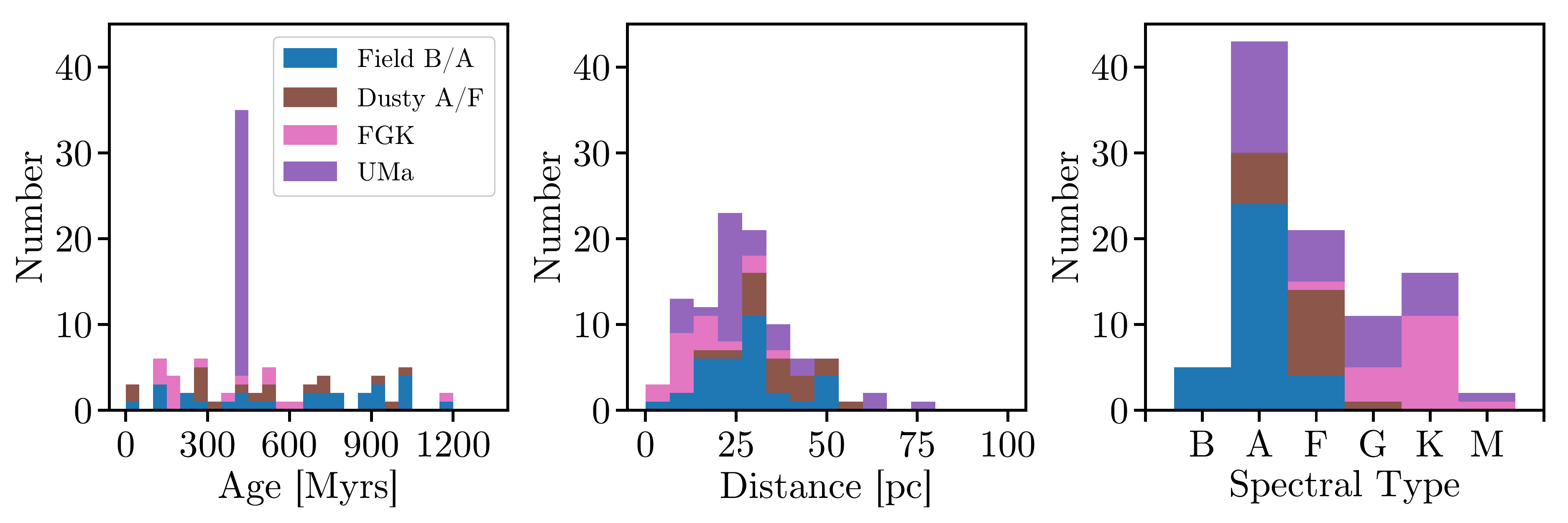}
\caption{Age, distance, and spectral-type distributions for our LEECH targets.
The large peak in the age distribution is due to our UMa subsample of similarly
aged stars.\label{histograms}} 
\end{figure*}

Figure \ref{histograms} provides a graphical summary of some of the most
relevant target parameters.  Our median target age is 400~Myr (driven by the
UMa sample), and our median distance is 25 pc. For comparison, the median target
in the \citet{Rameau2013} survey is 30 Myr at 40 pc, and the median target in
the International Deep Planet Survey \citep{Galicher2016} is 120 Myr at 45 pc.

\startlongtable
\begin{deluxetable*}{llllrrrcrrrc}
\tabletypesize{\footnotesize}
\tablecolumns{12}
\tablewidth{0pt}
\tablecaption{Target Summary\label{TargetSummary}}
\tablehead{
    \colhead{Name}& 
    \colhead{Name}&
    \colhead{R.A.}&
    \colhead{Decl.}&
    \colhead{Dist.}  &
    \colhead{$V$}&
    \colhead{$K$}&
    \colhead{SpT}&
    \colhead{$L^{\prime}$}&
    \colhead{Age\tablenotemark{a}}&
    \colhead{Mass}&
    \colhead{Sublist}\\
    \colhead{}&
    \colhead{}&
    \colhead{}&
    \colhead{}&
    \colhead{(pc)}&
    \colhead{(mag)}&
    \colhead{(mag)}&
    \colhead{}&
    \colhead{(mag)}&
    \colhead{(Myr)}&
    \colhead{($M_{\odot}$)}&
    \colhead{}}
\startdata
   HIP 544\tablenotemark{b} &         HD    166 &    00 06 36 &  +29 01 17 &  13.78 &   6.13 &   4.31     &      K0V &   4.25 & 155 &   0.95 & FGK \\
  HIP 1473                          &      $\sigma$ And &    00 18 19 &  +36 47 06 &  42.03 &   4.52 &   4.38     &      A2V &   4.37 & 450 &   2.10 & Dusty-A \\
  HIP 2843                          &         HD   3296 &    00 36 01 &  -05 34 14 &  44.90 &   6.71 &   5.59     &      F6V &   5.55 & 1000&   1.35 & Dusty-F \\
  HIP 6061                          &             f Psc &    01 17 47 &  +03 36 52 &  78.25 &   5.14 &   4.92     &      A1V &   4.92 & 412 &   2.52 & UMa \\
  HIP 8486\tablenotemark{c}         &            EZ Cet &    01 49 23 &  -10 42 13 &  23.67 &   6.75 &   5.15     &      G3V &   5.10 & 412 &   1.05 & UMa \\
  HIP 8497                          &        $\chi$ Cet &    01 49 35 &  -10 41 11 &  23.28 &   4.68 &   3.87     &      F0V &   3.84 & 412 &   1.55 & UMa \\
  HIP 8832\tablenotemark{d}        &      $\gamma$ Ari &    01 53 31 &  +19 17 37 &  50.05 &   4.52 &   4.7      &      A3m &   4.50 &  50 &   2.17 & Field-A \\
  HIP 8903\tablenotemark{c}         &       $\beta$ Ari &    01 54 38 &  +20 48 28 &  17.99 &   2.65 &   2.38     &      A5V &   2.36 & 931 &   2.08 & Field-A \\
 HIP 10064\tablenotemark{c}         &       $\beta$ Tri &    02 09 32 &  +34 59 14 &  38.90 &   3.00 &   2.66     &    A5III &   2.64 & 730 &   2.39 & Dusty-A \\
 HIP 10552\tablenotemark{c}         &         HD  13959 &    02 15 53 &  +06 37 34 &  35.93 &   9.07 &   6.31     &      K4V &   6.21 & 412 &   0.85 & UMa \\
 HIP 12828\tablenotemark{c}         &        $\mu$. Cet &    02 44 56 &  +10 06 50 &  26.50 &   4.20 &   3.46     &   A9III  &   3.43 & 1520&   1.73 & Field-A \\
 HIP 13402\tablenotemark{b} &         HD  17925 &    02 52 32 &  -12 46 10 &  10.36 &   6.05 &   4.06     &      K1V &   3.99 & 100 &   0.85 & FGK \\
 HIP 14576\tablenotemark{c}         &       $\beta$ Per &    03 08 10 &  +40 57 20 &  27.57 &   2.12 &   2.08     &      B8V &   2.08 & 447 &   3.48 & Field-B \\
 HIP 15457\tablenotemark{b} &         kap01 Cet &    03 19 21 &  +03 22 12 &   9.15 &   4.85 &   3.34     &      G5V &   3.29 & 400 &   1.00 & FGK \\
 HIP 16537\tablenotemark{b} &    $\epsilon$ Eri &    03 32 55 &  -09 27 29 &   3.20 &   3.73 &   1.67     &      K2V &   1.60 & 560 &   0.85 & FGK \\
 HIP 18859                          &         HD  25457 &     04 02 3 &   -00 16 0 &  18.77 &   5.38 &   4.18     &    F7/8V &   4.14 & 100 &   1.25 & FGK \\
 HIP 19859                          &         HD  26923 &    04 15 28 &  +06 11 12 &  22.08 &   6.30 &   4.90     &     G0IV &   4.85 & 412 &   1.10 & UMa \\
 HIP 19990                          &    $\omega$02 Tau &    04 17 15 &  +20 34 42 &  29.31 &   4.91 &   4.36     &      A3m &   4.35 & 1010&   1.65 & Field-A \\
 HIP 20901                          &             b Tau &    04 28 50 &  +13 02 51 &  48.57 &   5.01 &   4.53     &      A5m &   4.50 & 680 &   1.95 & Dusty-A \\
 HIP 21547                          &             c Eri &    04 37 36 &  -02 28 24 &  29.78 &   5.21 &   4.54     &     F0IV &   4.51 & 791 &   1.55 & Field-A \\
 HIP 22845\tablenotemark{b} &      $\pi$.01 Ori &    04 54 53 &  +10 09 02 &  34.23 &   4.65 &   4.42     &      A0V &   4.42 & 10  &   1.85 & Dusty-A \\
 HIP 23875                          &       $\beta$ Eri &    05 07 50 &  -05 05 11 &  27.40 &   2.79 &   2.38     &     A3IV &   2.37 & 791 &   2.60 & Field-A \\
 HIP 25428                          &       $\beta$ Tau &    05 26 17 &  +28 36 26 &  41.05 &   1.65 &   2.03     &    B7III &   2.03 & 122 &   4.07 & Field-B \\
 HIP 25486\tablenotemark{c}         &            AF Lep &    05 27 04 &  -11 54 03 &  26.87 &   6.30 &   4.93     &     F8V  &   4.88 & 40  &   1.20 & Dusty-F \\
 HIP 26779\tablenotemark{b} &         HD  37394 &    05 41 20 &  +53 28 51 &  12.28 &   6.23 &   4.27     &      K1V &   4.21 & 520 &   0.90 & FGK \\
 HIP 27072                          &      $\gamma$ Lep &    05 44 27 &  -22 26 54 &   8.88 &   3.60 &   2.41     &      F6V &   2.37 & 412 &   1.25 & UMa \\
 HIP 27913\tablenotemark{b}\tablenotemark{*} &      $\chi$01 Ori &    05 54 22 &  +20 16 34 &   8.84 &   4.40 &   2.99     &      G0V &   2.94 & 412 &   1.10 & UMa \\
 HIP 28360                          &       $\beta$ Aur &    05 59 31 &  +44 56 50 &  24.87 &   1.90 &   1.86     &   A1IV-V &   1.86 & 412 &   2.79 & UMa \\
 HIP 28954\tablenotemark{b} &         V1386 Ori &    06 06 40 &  +15 32 31 &  15.78 &   6.74 &   4.82     &      K0V &   4.76 & 600 &   0.90 & FGK \\
 HIP 33202                          &             e Gem &    06 54 38 &  +13 10 40 &  29.48 &   4.71 &    \nodata &      A8V &   4.11 & 1650&   1.60 & Field-A \\
 BD+201790                          &       BD+20  1790 &    07 23 43 &  +20 24 58 &  27.79 &  10.00 &   6.88     &       K5 &   6.77 & 180 &   0.70 & FGK \\
 HIP 36188                          &       $\beta$ CMi &    07 27 09 &  +08 17 21 &  49.14 &   2.89 &   3.03     &      B8V &   3.03 & 215 &   3.40 & Field-B \\
 HIP 36704                          &         HD  59747 &    07 33 00 &  +37 01 47 &  20.68 &   7.70 &   5.59     &      G5V &   5.54 & 412 &   0.85 & UMa \\
 HIP 41152                          &         HD  70313 &    08 23 48 &  +53 13 10 &  52.11 &   5.54 &   5.25     &      A3V &   5.24 & 250 &   1.90 & Dusty-A \\
 HIP 41820\tablenotemark{c}         &CCDM J08316+3458AB &    08 31 35 &  +34 57 58 &  27.32 &   7.51 &    \nodata &      G5V &   5.88 & 412 &   0.95 & UMa \\
 HIP 42438\tablenotemark{b} &      $\pi$.01 UMa &    08 39 11 &  +65 01 15 &  14.45 &   5.64 &   4.17     &    G1.5V &   4.12 & 412 &   1.05 & UMa \\
 HIP 43625                          &         HD  75616 &    08 53 06 &  +52 23 24 &  35.83 &   6.99 &   5.68     &       F5 &   5.64 & 400 &   1.20 & Dusty-F \\
 HIP 44127\tablenotemark{d}        &       $\iota$ UMa &    08 59 12 &  +48 02 30 &  14.16 &   3.14 &   2.66     &      A7V &   2.63 & 1010&   1.62 & Field-A \\
 HIP 44458\tablenotemark{d}        &         HD  77407 &    09 03 27 &  +37 50 27 &  30.21 &   7.04 &   5.44     &      G0V &   5.39 & 100 &   1.11 & FGK \\
  HD 78141                          &         HD  78141 &    09 07 18 &  +22 52 21 &  25.30 &   7.98 &   5.78     &       K0 &   5.72 & 150 &   0.90 & FGK \\
 HIP 44897                          &         HD  78366 &    09 08 51 &  +33 52 55 &  19.04 &   5.90 &   4.55     &   G0IV-V &   4.50 & 950 &   1.10 & Dusty-F \\
 HIP 44901                          &             f UMa &    09 08 52 &  +51 36 16 &  27.14 &   4.48 &   3.81     &     A3m  &   3.77 & 1400&   1.69 & Field-A \\
 HIP 46580                          &         HD  82106 &    09 29 54 &  +05 39 18 &  12.78 &   7.20 &   4.79     &      K3V &   4.70 & 370 &   0.80 & FGK \\
 HIP 48341                          &             6 Sex &    09 51 14 &  -04 14 36 &  65.23 &   6.02 &   5.64     &      A4V &   5.62 & 412 &   1.89 & UMa \\
 HIP 49593                          &            21 LMi &    10 07 25 &  +35 14 40 &  27.01 &   4.49 &   4.00     &   A7V    &   3.97 & 412 &   1.70 & UMa \\
 HIP 49669\tablenotemark{c}         &      $\alpha$ Leo &    10 08 22 &  +11 58 01 &  24.31 &   1.40 &   1.62     &    B8IV  &   1.62 & 253 &   3.80 & Field-B \\
 HIP 53910                          &       $\beta$ UMa &    11 01 50 &  +56 22 56 &  29.59 &   2.37 &   2.35     &     A1IV &   2.34 & 412 &   2.76 & UMa \\
 HIP 53985\tablenotemark{b} &            DS Leo &    11 02 38 &  +21 58 01 &  11.94 &   9.57 &   5.69     &      M1V &   5.48 & 412 &   0.55 & UMa \\
 HIP 54872                          &      $\delta$ Leo &    11 14 06 &  +20 31 25 &  17.91 &   2.53 &   2.24     &     A5IV &   2.22 & 858 &   2.05 & Field-A \\
 HIP 56997                          &            61 UMa &    11 41 03 &  +34 12 05 &   9.58 &   5.34 &   3.59     &      G8V &   3.53 & 1200&   0.94 & FGK \\
 HIP 57632                          &       $\beta$ Leo &    11 49 03 &  +14 34 19 &  11.00 &   2.13 &   1.90     &     A3Va &   1.89 & 412 &   1.95 & Field-A \\
 HIP 58001\tablenotemark{c}         &      $\gamma$ UMa &    11 53 49 &  +53 41 41 &  33.91 &   2.44 &   2.33     & A0V      &   2.33 & 412 &   2.82 & UMa \\
 HIP 58876                          &         HD 104860 &    12 04 33 &  +66 20 11 &  45.21 &   7.91 &   6.50     &       F8 &   6.46 & 250 &   1.10 & Dusty-F \\
 HIP 59774                          &      $\delta$ UMa &    12 15 25 &  +57 01 57 &  24.85 &   3.32 &   3.09     &      A2V &   3.08 & 412 &   2.19 & UMa \\
 HIP 61481                          &            DO CVn &    12 35 51 &  +51 13 17 &  26.25 &   8.54 &    \nodata &       K0 &   6.52 & 412 &   0.80 & UMa \\
 HIP 61946                          &         HD 110463 &    12 41 44 &  +55 43 28 &  22.73 &   8.28 &   6.00     &      K3V &   5.92 & 412 &   0.80 & UMa \\
 HIP 61960                          &        $\rho$ Vir &    12 41 53 &  +10 14 08 &  38.17 &   4.88 &   4.68     &      A0V &   4.68 & 500 &   1.80 & Dusty-A \\
 HIP 62512\tablenotemark{c}         &         HD 111456 &    12 48 39 &  +60 19 11 &  26.18 &   5.83 &   4.55     &      F6V &   4.51 & 412 &   1.30 & UMa \\
 HIP 62933                          &            41 Vir &    12 53 49 &  +12 25 06 &  60.72 &   6.24 &   5.47     &    A7III &   5.44 & 412 &   1.75 & UMa \\
 HIP 62956                          &    $\epsilon$ UMa &    12 54 01 &  +55 57 35 &  24.26 &   1.77 &   1.76     & A1III-IV &   1.76 & 412 &   2.79 & UMa \\
 HIP 63076                          &             8 Dra &    12 55 28 &  +65 26 18 &  29.39 &   5.22 &   4.43     &   A7m    &   4.40 & 250 &   1.52 & Dusty-F \\
 HIP 63125                          &    $\alpha$02 CVn &    12 56 01 &  +38 19 06 &  37.64 &   2.88 &   3.16     & A0V      &   3.16 & 112 &   2.71 & Field-B \\
 HIP 63503\tablenotemark{d}        &            78 UMa &    13 00 43 &  +56 21 58 &  25.52 &   4.93 &   3.95     &      F2V &   3.92 & 412 &   1.55 & UMa \\
 HIP 65327                          &         HD 238224 &    13 23 23 &  +57 54 22 &  24.11 &   9.56 &   6.42     &      K5V &   6.31 & 412 &   0.70 & UMa \\
 HIP 65378\tablenotemark{c}         &       $\zeta$ UMa &    13 23 55 &  +54 55 31 &  26.31 &   2.27 &    \nodata &  A1.5V   &   2.16 & 412 &   2.67 & UMa \\
 HIP 66459\tablenotemark{b} &       BD+36  2393 &    13 37 28 &  +35 43 03 &  10.96 &   9.07 &   5.49     &   K7.5V  &   5.35 & 412 &   0.60 & UMa \\
 HIP 69713                          &       $\iota$ Boo &    14 16 09 &  +51 22 02 &  29.15 &   4.75 &   4.29     &      A7V &   4.26 & 1010&   1.70 & Field-A \\
 HIP 69732                          &     $\lambda$ Boo &    14 16 23 &  +46 05 17 &  30.34 &   4.18 &   3.91     &  A0V     &   3.91 & 290 &   2.00 & Dusty-A \\
 HIP 69989                          &            18 Boo &    14 19 16 &  +13 00 15 &  26.23 &   5.40 &   4.39     &     F5IV &   4.35 & 412 &   1.40 & UMa \\
 HIP 71075                          &      $\gamma$ Boo &    14 32 04 &  +38 18 29 &  26.74 &   3.02 &   2.56     & A7IV     &   2.53 & 858 &   2.03 & Field-A \\
 HIP 71876                          &            DL Dra &    14 42 03 &  +61 15 42 &  41.24 &   6.25 &    \nodata &      F2V &   5.40 & 412 &   1.40 & UMa \\
 HIP 72603\tablenotemark{c}         &    $\alpha$01 Lib &    14 50 41 &  -15 59 50 &  23.37 &   5.16 &   4.14     &      F3V &   4.11 & 300 &   1.40 & Dusty-F \\
 HIP 72659\tablenotemark{d}        &        $\xi$ Boo  &    14 51 23 &  +19 06 01 &   6.73 &   4.59 &   1.97     & G7V      &   1.92 & 290 &   1.00 & FGK \\
 HIP 72848\tablenotemark{c}         &            DE Boo &    14 53 23 &  +19 09 10 &  11.38 &   6.01 &   4.32     &    K0.5V &   4.26 & 1300&   0.90 & FGK \\
 HIP 74702                          &         HD 135599 &    15 15 59 &  +00 47 46 &  15.82 &   6.91 &   4.96     &      K0V &   4.90 & 1300&   0.85 & FGK \\
 HIP 76267\tablenotemark{c}         &      $\alpha$ CrB &    15 34 41 &  +26 42 52 &  23.01 &   2.24 &   2.21     &     A1IV &   2.21 & 412 &   2.60 & UMa \\
 HIP 77233                          &       $\beta$ Ser &    15 46 11 &  +15 25 18 &  46.66 &   3.67 &   3.42     &     A2IV &   3.41 & 412 &   2.67 & UMa \\
 HIP 77622                          &    $\epsilon$ Ser &    15 50 48 &  +04 28 39 &  20.80 &   3.69 &   3.42     &     A5m  &   3.40 & 672 &   1.75 & Field-A \\
 HIP 80459\tablenotemark{b} &          G 202-48 &    16 25 24 &  +54 18 14 &   6.47 &  10.17 &   5.83     &    M1.5V &   5.61 & 500 &   0.40 & FGK \\
 HIP 83207\tablenotemark{c}         &    $\epsilon$ Her &    17 00 17 &  +30 55 35 &  50.08 &   3.92 &   3.92     &      A0V &   3.92 & 215 &   2.91 & Field-B \\
 HIP 84379                          &      $\delta$ Her &    17 15 01 &  +24 50 21 &  23.33 &   3.13 &   2.85     & A1IV     &   2.84 & 729 &   2.25 & Field-A \\
 HIP 85829\tablenotemark{c}         &      $\nu$.02 Dra &    17 32 16 &  +55 10 22 &  29.98 &   4.83 &   4.16     &    A4m   &   4.13 & 700 &   1.62 & Dusty-A \\
 HIP 86032\tablenotemark{c}         &      $\alpha$ Oph &    17 34 56 &  +12 33 36 &  14.90 &   2.07 &   1.66     &    A5III &   1.64 & 931 &   2.10 & Field-A \\
 HIP 88771                          &            72 Oph &    18 07 20 &  +09 33 49 &  26.81 &   3.73 &   3.42     &      A5V &   3.40 & 931 &   2.00 & Field-A \\
 HIP 91262                          &      $\alpha$ Lyr &    18 36 56 &  +38 47 01 &   7.68 &   0.03 &   0.13     &     A0V  &   0.13 & 485 &   2.41 & Field-A \\
 HIP 92161                          &           111 Her &    18 47 01 &  +18 10 53 &  28.29 &   4.36 &   4.08     &    A5III &   4.06 & 672 &   1.85 & Field-A \\
 HIP 93408                          &            16 Lyr &    19 01 26 &  +46 56 05 &  38.61 &   5.01 &   4.51     &     A6IV &   4.48 & 412 &   1.85 & UMa \\
 HIP 93747                          &       $\zeta$ Aql &    19 05 24 &  +13 51 48 &  23.52 &   2.99 &   2.88     & A0IV-V   &   2.88 & 526 &   2.60 & Field-A \\
 HIP 93805                          &     $\lambda$ Aql &    19 06 14 &  -04 52 57 &  36.95 &   3.43 &   3.65     &     B9V  &   3.65 & 122 &   2.66 & Field-B \\
 HIP 97165\tablenotemark{d}        &      $\delta$ Cyg &    19 44 58 &  +45 07 50 &  51.26 &   2.87 &   2.83     &     A0IV &   2.83 & 350 &   3.14 & Field-B \\
 HIP 97649                          &      $\alpha$ Aql &    19 50 46 &  +08 52 05 &   5.13 &   0.76 &   0.24     &     A7V  &   0.21 & 1010&   1.70 & Field-A \\
HIP 105199                          &      $\alpha$ Cep &    21 18 34 &  +62 35 08 &  15.04 &   2.46 &   2.07     &     A8V  &   2.04 & 1190&   2.20 & Field-A \\
HIP 105918                          &         HD 204277 &    21 27 06 &  +16 07 26 &  33.11 &   6.72 &   5.45     &      F8V &   5.41 & 900 &   1.18 & Dusty-F \\
HIP 107556\tablenotemark{c}         &      $\delta$ Cap &    21 47 02 &  -16 07 38 &  13.63 &   2.83 &   2.14     &      A5m &   2.11 & 1400&   1.61 & Field-A \\
HIP 109427                          &      $\theta$ Peg &    22 10 11 &  +06 11 52 &  27.20 &   3.55 &   3.33     &     A1V  &   3.32 & 729 &   2.10 & Field-A \\
 BD+483686                          &          V383 Lac &    22 20 07 &  +49 30 11 &  34.44 &   8.58 &   6.51     &      K1V &   6.44 & 150 &   0.90 & FGK \\
HIP 111278                          &            39 Peg &    22 32 35 &  +20 13 48 &  53.59 &   6.44 &   5.58     &      F1V &   5.55 & 500 &   1.55 & Dusty-F \\
HIP 114570                          &             7 And &    23 12 33 &  +49 24 22 &  24.40 &   4.52 &   3.77     &      F1V &   3.74 & 1400&   1.56 & Field-A \\
\enddata 
\tablenotetext{a}{Stellar ages for stars in the FGK subsample are from
\citet{Mamajek2008} and \citet{Heinze2010A}. Ages for stars in the Dusty-A/F
subsample are from \citet{Gaspar2013}. Ages for stars in the UMa subsample are
from \citet{Jones2015}. Ages for stars in the Field-B/A subsample are derived
in this work.} 
\tablenotetext{b}{Our photometric sensitivity was sufficient to detect
$\leq10~M_{\mathrm{Jup}}$ cold-start planets in these systems.}
\tablenotetext{c}{Close ($\lesssim1\arcsec$) binary systems}
\tablenotetext{d}{Wide ($\gtrsim1\arcsec$) binary systems}
\end{deluxetable*}

\subsection{Age Determination for the Field-B/A Sublist}\label{fieldBAsubsec}
We constructed the Field-B/A sublist by querying the union of the Tycho2 and
$Hipparcos$ catalogs for stars observable from the LBT ($\delta>-20^{\circ}$)
with $B-V<0.33$. We required target distances to be less than 30 pc for stars
with $B-V\geq0$ (A-type stars) and less than 55 pc for stars with $B-V<0$
(B-type stars), accommodating the relatively low frequency of the more massive
type.  Two white dwarfs and all of the objects with pre-\textit{Gaia} distance
measurements uncertain at $\geq5\%$ were removed. 

We closely followed the work of \citet{Nielsen2013} to determine an age for
each observed target in the Field-B/A sublist, incorporating knowledge of the
local stellar population to implement a Bayesian approach to derive a posterior
distribution function of age ($t_{*}$), mass ($M_{*}$), and the log of the
metallicity ratio with respect to the Sun ($z_{*}$).

Explicitly, we calculated a likelihood function for each target according to 
\begin{equation}
P(V,(B-V)|t_{*},M_{*},z_{*}) \propto e^{-\frac{1}{2}\chi^{2}(t_{*},M_{*},z_{*})}
\end{equation}
with
\begin{equation}\label{chieq}
\chi^{2}(t_{*},M_{*},z_{*}) = \frac{(V - V_{\mathrm{m}})^{2}}{\sigma_{V}^{2}} + \frac{((B-V) - (B-V)_{\mathrm{m}})^{2}}{\sigma_{(B-V)}^2}.
\end{equation}
In equation \ref{chieq}, $V$ and $(B-V)$ are, respectively, the observed
absolute $V$-band magnitude and the $B-V$ color from the extended $Hipparcos$
catalog \citep{Anderson2012}. Both $V_{\mathrm{m}}$ and $(B-V)_{m}$ are
functions of $t_{*}$, $M_{*}$, and $z_{*}$, and are the model predictions for
absolute $V$-band magnitude and $B-V$ color from the PARSEC isochrones
\citep{Marigo2017}. Since the extended $Hipparcos$ compilation does not include
an uncertainty entry for the absolute $V$-band magnitude, we estimated
$\sigma_{V}$ by combining a propagation of the distance uncertainty with
$1/\sqrt{2}$ times $\sigma_{(B-V)}$, the error on the measured $B-V$
color. 

To derive our posterior distribution, we used a prior that is uncorrelated
between the model parameters, 
\begin{equation}
P(t_{*},M_{*},z_{*}) = P(t_{*})P(M_{*})P(z_{*}).  
\end{equation}
We used a uniform prior in linear age. This implies a constant star-formation
rate over an age range that includes all of our targets, which is consistent
with the results of \citet{Cignoni2006}. Like in \citet{Nielsen2013}, our
isochrone grid is logarithmic in age. Our prior accounts for the different
linear age intervals represented by each grid point with the effect of pushing
fits to older ages, where grid steps are wider. For the mass, we assumed
a Salpeter initial mass function $dN/dM \propto M^{-2.35}$
\citep{Salpeter1955}, with the effect of pushing fits to lower masses.
Finally, for metallicity, we were again guided by \citet{Nielsen2013}  and
approximated the results of \citet{Casagrande2011}  and \citet{Nieva2012} by
taking the metallicity distribution of nearby stars to be Gaussian distributed
with mean -0.05 and $\sigma$ 0.11 dex. 

In Table \ref{FieldABtable}, we summarize our Bayesian fitting results.  For
each parameter, we report the peak of the marginalized posterior probability
distribution as the best fit. We also report the parameter ranges corresponding
to the intervals between 16\% and 84\% and between 2.5\% and 97.5\% in the
cumulative distribution function of the posterior. 

One of the targets in our Field-B/A sublist, HIP~25428, is actually
a high-confidence (96\%) member of the AB Dor moving group based on its
kinematics according to the online BANYAN $\Sigma$ Bayesian analysis tool
\citep{Gagne2018}. The AB~Dor moving group has an age of $149_{-19}^{+51}$~Myr
\citep{Bell2015}.  We derive an age of 122 Myr for HIP 25428 with with 95\%
confidence range due to statistical uncertainty ranging from 104 to 124 Myr. We
show below that significant systematic uncertainty ($\sim50\%$) is also
involved in age-dating stars with model isochrone fitting.  Thus, we take our
fitted value as validation that our approach returns sensible results. Another
target, HIP~57632 ($\beta$~Leo), is often reported to have an age of $\sim40$~Myr
based on assumed membership in either IC 2391 \citep{Eggen1991} or the Argus
Association \citep{Zuckerman2011}. Our best-fit age of 412~Myr is in some
tension with the younger age.  However, other authors also derive an older age
for HIP~57632: \citet{Nielsen2013} find a 95\% confidence interval of
16-458~Myr, and \citet{Rieke2005} find a best-fit age of 520~Myr.  While our
age determination for HIP~57632 is highly uncertain (56-685~Myr, 95\%
confidence), the older ages may not be incorrect. The BANYAN $\Sigma$ tool
reports 0\% probability of HIP~57632 membership in IC 2391 and leaves out an
analysis of the Argus Association as this may not be a co-eval group of stars
\citep{Bell2015}.  HIP~57632 hosts a debris disk, but, as shown by
\citet{Rieke2005}, the 24~$\mu$m flux excess is not inconsistent with the upper
envelope of debris-disk luminosities for an age of 520~Myr.

\begin{figure} 
\plotone{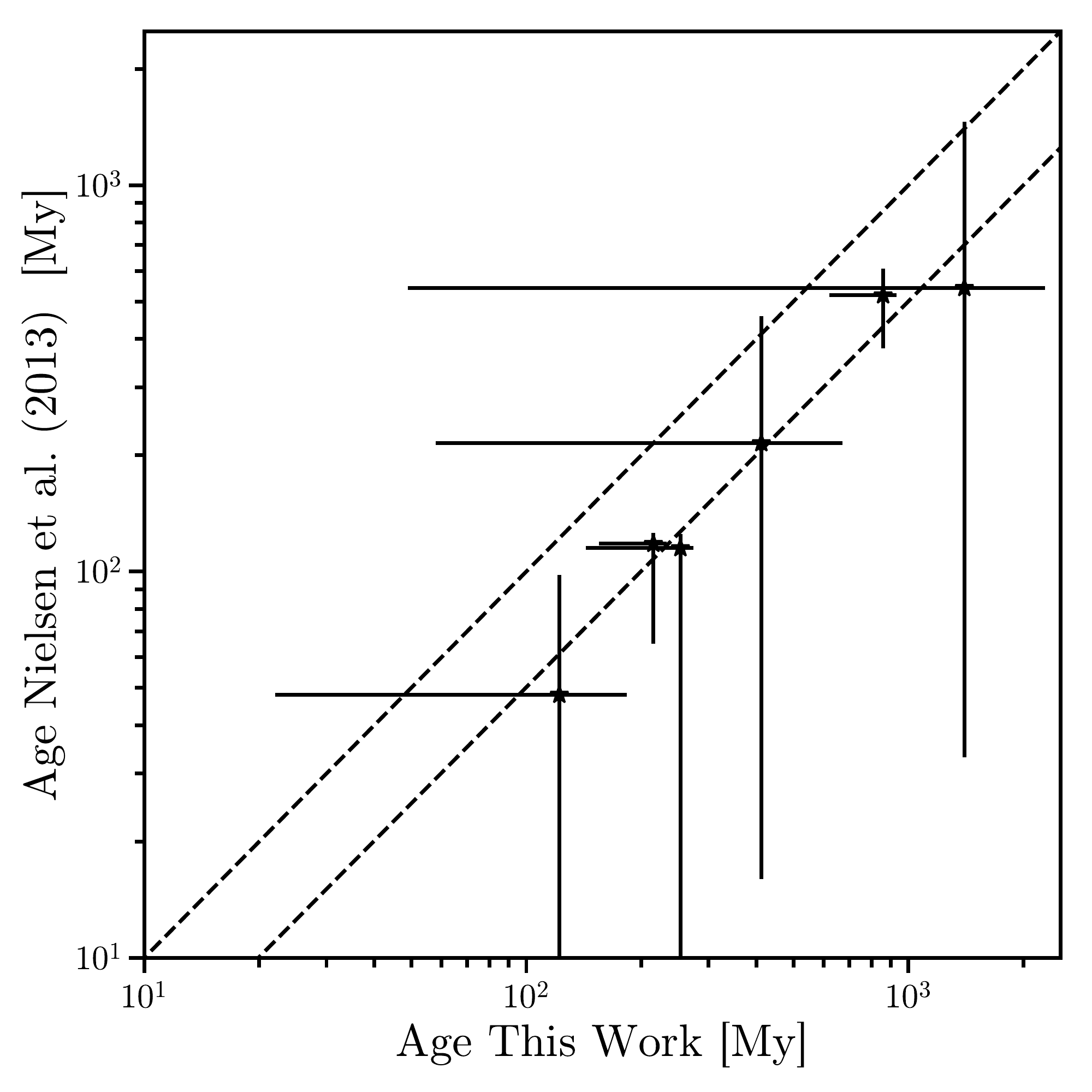} 
\caption{Ages derived here versus ages derived by \citet{Nielsen2013} for the
six targets shared by both surveys. From left to right, the target stars are
HIP~93805, HIP~36188, HIP~49669, HIP~57632, HIP~54872, and HIP~107556. The
upper dashed line indicates a one-to-one relationship, the lower dashed line
one-to-two. Our age estimates are systematically older than those found by
\citet{Nielsen2013}, suggesting the magnitude of systematic error inherent to
the model isochrones.\label{AgeFig}} 
\end{figure}

In all, we share six targets with \citet{Nielsen2013}, who used the
\citet{Siess2000} isochrones to perform their fits. We plot our derived ages
versus the \citet{Nielsen2013} ages in Figure \ref{AgeFig}. Error bars show the
corresponding 95\%-confidence intervals. From left to right, the objects are 
HIP~93805, HIP~36188, HIP~49669, HIP~57632, HIP~54872, and HIP~107556.
The ages we derive are systematically older than the ages derived by
\citet{Nielsen2013}.  For the stars common to both studies, the
\citet{Nielsen2013} ages are 50\% younger than LEECH ages, on average. This
suggests the magnitude of systematic error inherent in the model isochrones.
The different inferred ages may be due to differences in the treatment of
mixing within stars (e.g., overshoot in the convective core), which can prolong
main-sequence lifetimes. By pushing targets to older ages, our models are more
conservative because we have less sensitivity to planets around older stars.

\begin{deluxetable*}{llllllllllllllll}
\tabletypesize{\scriptsize}
\tablecolumns{16}
\tablewidth{0pt}
\tablecaption{Best-fit Age, Mass, and Metallicity for observed stars in the Field-B/A sublist\label{FieldABtable}}
\tablehead{
    \colhead{Name}&
    \multicolumn{5}{c}{Age [Myr]}&
    \multicolumn{5}{c}{Mass [$M_{\odot}$]}&
    \multicolumn{5}{c}{metallicity [$\log_{10}(z/z_{\odot})$]} \\
    \colhead{}&
    \colhead{Mode}&
    \multicolumn{2}{c}{16\%-84\%}&
    \multicolumn{2}{c}{2.5\%-97.5\%}&
    \colhead{Mode}&
    \multicolumn{2}{c}{16\%-84\%}&
    \multicolumn{2}{c}{2.5\%-97.5\%}&
    \colhead{Mode}&
    \multicolumn{2}{c}{16\%-84\%}&
    \multicolumn{2}{c}{2.5\%-97.5\%}}
\startdata
HIP   8832  &  50   & 14   & 81   & 7    & 151  & 2.2 & 2.1 & 2.3 & 2.1 & 2.3 & -0.26 & -0.27 & -0.13 & -0.30 & -0.04 \\    
HIP   8903  &  931  & 808  & 989  & 704  & 5477 & 1.9 & 1.9 & 2.0 & 1.8 & 2.0 & -0.04 & -0.14 &  0.05 & -0.25 &  0.18 \\     
HIP  12828  &  1517 & 1249 & 1629 & 1045 & 1773 & 1.6 & 1.5 & 1.6 & 1.5 & 1.7 & -0.05 & -0.15 &  0.06 & -0.23 &  0.17 \\     
HIP  14576  &  447  & 389  & 438  & 373  & 465  & 2.8 & 2.7 & 2.8 & 2.6 & 2.9 &  0.03 & -0.15 &  0.05 & -0.24 &  0.17 \\     
HIP  19990  &  1010 & 403  & 1095 & 108  & 1358 & 1.5 & 1.5 & 1.6 & 1.4 & 1.6 & -0.03 & -0.15 &  0.06 & -0.24 &  0.13 \\     
HIP  21547  &  791  & 190  & 937  & 44   & 1265 & 1.5 & 1.4 & 1.5 & 1.4 & 1.5 & -0.04 & -0.16 & -0.01 & -0.25 &  0.04 \\     
HIP  23875  &  791  & 689  & 776  & 662  & 820  & 2.2 & 2.2 & 2.3 & 2.1 & 2.3 & -0.10 & -0.11 &  0.05 & -0.27 &  0.19 \\     
HIP  25428  &  122  & 106  & 119  & 104  & 124  & 4.5 & 4.4 & 4.5 & 4.3 & 4.5 & -0.14 & -0.19 &  0.07 & -0.27 &  0.17 \\     
HIP  33202  &  1646 & 1159 & 1723 & 834  & 1919 & 1.5 & 1.5 & 1.6 & 1.4 & 1.6 & -0.05 & -0.15 &  0.06 & -0.24 &  0.17 \\     
HIP  36188  &  215  & 174  & 223  & 151  & 247  & 3.5 & 3.3 & 3.6 & 3.2 & 3.7 & -0.05 & -0.15 &  0.06 & -0.25 &  0.17 \\     
HIP  44127  &  1010 & 606  & 1086 & 346  & 1284 & 1.6 & 1.6 & 1.7 & 1.5 & 1.7 & -0.06 & -0.15 &  0.06 & -0.24 &  0.17 \\     
HIP  44901  &  1399 & 1131 & 1516 & 932  & 1686 & 1.6 & 1.6 & 1.7 & 1.5 & 1.7 & -0.03 & -0.15 &  0.06 & -0.25 &  0.17 \\     
HIP  49669  &  253  & 179  & 255  & 138  & 287  & 3.3 & 3.2 & 3.5 & 3.1 & 3.8 & -0.03 & -0.15 &  0.06 & -0.25 &  0.17 \\     
HIP  54872  &  858  & 704  & 898  & 606  & 982  & 1.9 & 1.9 & 2.0 & 1.8 & 2.1 & -0.05 & -0.15 &  0.07 & -0.25 &  0.17 \\     
HIP  57632  &  412  & 200  & 530  & 56   & 685  & 1.8 & 1.8 & 1.9 & 1.7 & 1.9 & -0.11 & -0.16 &  0.05 & -0.26 &  0.13 \\     
HIP  62956  &  380  & 357  & 399  & 345  & 411  & 2.9 & 2.7 & 2.9 & 2.7 & 3.0 &  0.02 & -0.15 &  0.03 & -0.17 &  0.19 \\     
HIP  63076  &  1010 & 319  & 1160 & 72   & 1483 & 1.5 & 1.4 & 1.5 & 1.4 & 1.5 & -0.06 & -0.15 &  0.04 & -0.24 &  0.09 \\     
HIP  63125  &  112  & 47   & 126  & 19   & 166  & 3.4 & 3.2 & 3.5 & 3.1 & 3.7 & -0.05 & -0.16 &  0.05 & -0.25 &  0.15 \\     
HIP  69713  &  1010 & 492  & 1102 & 169  & 1357 & 1.6 & 1.5 & 1.6 & 1.5 & 1.7 & -0.04 & -0.15 &  0.06 & -0.24 &  0.16 \\     
HIP  71075  &  858  & 799  & 903  & 732  & 927  & 2.1 & 2.0 & 2.1 & 2.0 & 2.2 &  0.02 & -0.15 &  0.03 & -0.17 &  0.17 \\     
HIP  77622  &  672  & 391  & 782  & 190  & 952  & 1.7 & 1.7 & 1.8 & 1.6 & 1.9 & -0.05 & -0.15 &  0.06 & -0.25 &  0.17 \\     
HIP  83207  &  447  & 389  & 477  & 341  & 516  & 2.5 & 2.4 & 2.5 & 2.3 & 2.6 & -0.02 & -0.14 &  0.06 & -0.25 &  0.17 \\     
HIP  84379  &  729  & 565  & 759  & 464  & 835  & 2.0 & 1.9 & 2.1 & 1.9 & 2.1 & -0.08 & -0.15 &  0.06 & -0.23 &  0.16 \\     
HIP  86032  &  931  & 781  & 952  & 690  & 1468 & 1.9 & 1.9 & 2.0 & 1.8 & 2.1 & -0.08 & -0.17 &  0.09 & -0.28 &  0.17 \\     
HIP  88771  &  931  & 755  & 991  & 633  & 1079 & 1.8 & 1.8 & 1.9 & 1.8 & 2.0 & -0.06 & -0.14 &  0.06 & -0.23 &  0.16 \\     
HIP  91262  &  485  & 431  & 525  & 380  & 565  & 2.4 & 2.3 & 2.4 & 2.3 & 2.5 &  0.01 & -0.16 &  0.07 & -0.24 &  0.17 \\     
HIP  92161  &  672  & 408  & 785  & 210  & 944  & 1.7 & 1.7 & 1.8 & 1.6 & 1.9 & -0.04 & -0.15 &  0.06 & -0.25 &  0.17 \\     
HIP  93747  &  526  & 428  & 563  & 357  & 616  & 2.2 & 2.2 & 2.3 & 2.1 & 2.3 & -0.05 & -0.14 &  0.06 & -0.23 &  0.16 \\     
HIP  93805  &  122  & 54   & 143  & 21   & 188  & 3.0 & 2.9 & 3.1 & 2.8 & 3.2 & -0.06 & -0.16 &  0.05 & -0.25 &  0.15 \\     
HIP  97165  &  350  & 247  & 345  & 235  & 350  & 3.1 & 3.1 & 3.3 & 3.0 & 3.4 &  0.04 & -0.15 &  0.07 & -0.27 &  0.12 \\     
HIP  97649  &  1010 & 690  & 1141 & 454  & 1330 & 1.6 & 1.6 & 1.7 & 1.5 & 1.8 & -0.05 & -0.15 &  0.06 & -0.24 &  0.17 \\     
HIP 105199  &  1188 & 1045 & 1244 & 942  & 2358 & 1.8 & 1.8 & 1.9 & 1.7 & 1.9 & -0.04 & -0.16 &  0.06 & -0.28 &  0.16 \\     
HIP 107556  &  1399 & 256  & 1633 & 47   & 2280 & 1.6 & 1.5 & 1.6 & 1.4 & 1.7 & -0.06 & -0.17 &  0.04 & -0.26 &  0.14 \\     
HIP 109427  &  729  & 594  & 775  & 500  & 845  & 2.0 & 2.0 & 2.1 & 1.9 & 2.1 & -0.11 & -0.12 &  0.07 & -0.22 &  0.17 \\     
HIP 114570  &  1399 & 927  & 1553 & 571  & 1778 & 1.5 & 1.5 & 1.6 & 1.4 & 1.6 & -0.05 & -0.15 &  0.07 & -0.24 &  0.17 \\ 
\enddata 
\end{deluxetable*}

\section{Observations} \label{Observations} Our survey was conducted using the
LBTI instrument \citep{Hinz2016} at the LBT on Mt.  Graham in southern Arizona.
LBTI is located between the two 8.4 m primary mirrors of the LBT at the
combined bent Gregorian focus. Light from each side of the telescope is
corrected for atmospheric aberrations using the LBTI AO system
\citep{Bailey2014} and delivered into the instrument via a cryogenic beam
combiner where it is then directed to individual science modules.  For LEECH
observations, we used the LMIRcam module of LBTI, which is optimized for work
in the thermal-infrared \citep[$3-5$ $\mu$m;][]{Skrutskie2010,Leisenring2012}.
LBTI does not include an instrument derotator, so images always rotate with
respect to the detector pixels as the parallactic angle changes.

During the course of the LEECH survey, LMIRcam provided an
$11\arcsec\times11\arcsec$ field of view, reading a $1024\times1024$ subsection
of its 5.2 $\mu$m cutoff HAWAII-2RG detector (the full $2048\times2048$ extent
of the array now provides a $20\arcsec\times20\arcsec$ field of view with
LMIRcam). LMIRcam was designed with a plate scale to accommodate imaging
interferometry at the full resolution of the 23~m LBT (10.7 mas
pixel$^{-1}$).  However, for LEECH observations, we operated without overlapping and
interfering the beams of the two primary mirrors, opting to make two images of
each source on the detector instead. In this mode, the $L^{\prime}$ images from
each side were oversampled, providing added robustness to bad pixels and cosmic
rays.

Our simultaneous non-overlapped imaging strategy provided increased sensitivity
in speckle-limited regimes because each image displays a mostly independent
speckle pattern. However, in the background-limited regime, our sensitivity was
decreased because of the presence of two sky backgrounds per image. 

Our standard observing procedure aligned both images of a target star in the
upper section of the detector keeping each image $3\arcsec$ from the edge of
the detector and leaving $5\arcsec$ between the images. For some very nearby
targets (e.g., Vega, Altair), we used only one side of the telescope and created
only one centered image in order to maximize our field of view. We also
performed single-sided observations during intervals of technical downtime
affecting either of the AO systems. In particular, all
observations conducted during the spring of 2013 were made with the left side
of the telescope only. 

We collected data using an up-down nod pattern (left-right for some of our
single-sided observations of binary systems) to track low spatial frequency
variations in the sky background level and high  and low spatial frequency
detector drifts.  A typical sequence included 50 frames in each position each
composed of three coadds of 0.3~s exposures. Our exposure time was set to
balance image saturation and detector readout efficiency.  Our choice of
exposure time generally resulted in a saturated core in each of our stellar
images. (We do not use a coronagraph for LEECH observations.) Our nod frequency
was set to optimize on-sky efficiency given the demands of a variable
background and detector.

An observing log for the LEECH survey is presented in Table \ref{ObservingLog}.
Differences in exposure time and rotation angle between the two sides of LBT
can result from differences in the AO performance (poor Strehl/open AO loop) that
cause some frames to be rejected for one side but not the other (see Section
\ref{DataProcessing}). 

\startlongtable
\begin{deluxetable}{lllrrrr}
\tabletypesize{\tiny}
\tablecolumns{7}
\tablewidth{0pt}
\tablecaption{LEECH Observing Log\label{ObservingLog}}
\tablehead{
           &
           &
           &
           &
           &
           &
                  \\
                  &
                  &
                  &
    \multicolumn{2}{c}{Left Side}&
    \multicolumn{2}{c}{Right Side}\\
    \colhead{Name}& 
    \colhead{Date}&
    \colhead{Seeing\tablenotemark{a}}  &
    \colhead{$t_{\mathrm{int}}$}&
    \colhead{Rot.}&
    \colhead{$t_{\mathrm{int}}$}&
    \colhead{Rot.}\\
    \colhead{}&
    \colhead{}&
    \colhead{($\arcsec$)}&
    \colhead{(s)}&
    \colhead{($^\circ$)}&
    \colhead{(s)}&
    \colhead{($^\circ$)}}
\startdata
HIP  54872  &  2013 Apr 18  &  $1.7\pm0.2$  & 2043  & 67     & \nodata  & \nodata\\
HIP  69713  &  2013 Apr 18  &  $2.0\pm0.2$  & 1638  & 53     & \nodata  & \nodata\\
HIP  86032  &  2013 Apr 18  &  $1.4\pm0.2$  & 1716  & 17     & \nodata  & \nodata\\
HIP  88771  &  2013 Apr 19  &   1.2        & 1256  & 17     & \nodata  & \nodata\\
HIP  77622  &  2013 Apr 19  &   1.2        & 10    &  0     & \nodata  & \nodata\\
HIP  76267  &  2013 Apr 20  &  $0.9\pm0.1$  & 1979  & 107    & \nodata  & \nodata\\
HIP  66459  &  2013 Apr 20  &  $0.9\pm0.1$  & 2440  & 161    & \nodata  & \nodata\\
HIP  42438  &  2013 Apr 21  &  $0.9\pm0.2$  & 4199  & 48     & \nodata  & \nodata\\
HIP  84379  &  2013 Apr 21  &  $0.9\pm0.1$  & 3153  & 96     & \nodata  & \nodata\\
HIP  46580  &  2013 Apr 22  &  $1.1\pm0.2$  & 3276  & 51     & \nodata  & \nodata\\
HIP  72659  &  2013 Apr 22  &  $0.9\pm0.1$  & 1909  & 45     & \nodata  & \nodata\\
HIP  62512  &  2013 Apr 22  &  $1.2\pm0.2$  & 3197  & 38     & \nodata  & \nodata\\
HIP  49593  &  2013 Apr 23  &   1.0        & 3271  & 158    & \nodata  & \nodata\\
HIP  93408  &  2013 Apr 23  &  $1.0\pm0.1$  & 975   & 6      & \nodata  & \nodata\\
HIP  61946  &  2013 Apr 23  &  $0.9\pm0.1$  & 4738  & 68     & \nodata  & \nodata\\
HIP  83207  &  2013 Apr 24  &   0.8        & 2635  & 150    & \nodata  & \nodata\\
HIP  57632  &  2013 Apr 24  &  $1.2\pm0.2$  & 3333  & 61     & \nodata  & \nodata\\
HIP  72659  &  2013 Apr 24  &  $0.8\pm0.1$  & 1965  & 34     & \nodata  & \nodata\\
HIP  62512  &  2013 Apr 24  &  $0.9\pm0.1$  & 1465  & 14     & \nodata  & \nodata\\
HIP  65378  &  2013 Apr 26  &  $1.1\pm0.2$  & \nodata   & \nodata    & 4599 & 63\\
HIP  77233  &  2013 Apr 26  &  $0.8\pm0.1$  & 3872  & 44     & \nodata  & \nodata\\
HIP  92161  &  2013 Apr 26  &   0.7        & \nodata   & \nodata    & 4124 & 58\\
HIP  69713  &  2013 May 23  &  $0.8\pm0.0$  & 3073  & 96     & \nodata  & \nodata\\
HIP  93408  &  2013 May 23  &   0.7        & 175   & 0      & \nodata  & \nodata\\
HIP  97649  &  2013 May 24  &  $0.9\pm0.2$  & 2873  & 83     & \nodata  & \nodata\\
HIP  69989  &  2013 May 25  &  $0.9\pm0.1$  & 2468  & 75     & \nodata  & \nodata\\
HIP  77622  &  2013 May 25  &  $0.9\pm0.1$  & 2285  & 47     & \nodata  & \nodata\\
HIP  93747  &  2013 Jun 17  &  $1.0\pm0.1$  & 1220  & 66     & \nodata  & \nodata\\
HIP  72659  &  2013 Jun 17  &  $0.9\pm0.1$  & 3012  & 82     & \nodata  & \nodata\\
HIP  71876  &  2013 Jun 26  &  $0.8\pm0.1$  & 1611  & 47     & \nodata  & \nodata\\
HIP  72848  &  2013 Jun 27  &  $0.9\pm0.1$  & 2708  & 69     & \nodata  & \nodata\\
HIP  91262  &  2013 Jun 27  &   0.6        & \nodata   & \nodata    & \nodata  & \nodata\\
HIP 105199  &  2013 Oct 18  &  $1.1\pm0.1$  & 2534  & 42     & 2552 & 47\\
HIP    544  &  2013 Oct 18  &  $1.0\pm0.1$  & 1475  & 133    & 1465 & 133\\
HIP  27072  &  2013 Oct 18  &  $1.0\pm0.1$  & 846   & 15     & 859  & 15\\
HIP  27913  &  2013 Oct 20  &  $0.9\pm0.1$  & \nodata   & \nodata    & 2054 & 44\\
HIP  18859  &  2013 Oct 21  &  $0.8\pm0.1$  & \nodata   & \nodata    & 441  & 9\\
HIP  28954  &  2013 Oct 21  &  $1.0\pm0.1$  & 4078  & 89     & 4037 & 82\\
HIP   8903  &  2013 Oct 22  &  $1.1\pm0.2$  & \nodata   & \nodata    & 2374 & 75\\
HIP  19990  &  2013 Oct 23  &  $1.0\pm0.1$  & \nodata   & \nodata    & 6675 & 117\\
HIP  26779  &  2013 Oct 24  &  $0.9\pm0.1$  & \nodata   & \nodata    & 6048 & 117\\
HIP  14576  &  2013 Oct 24  &  $0.9\pm0.1$  & \nodata   & \nodata    & 1652 & 51\\
HIP  42438  &  2013 Nov 19  &  $1.4\pm0.2$  & \nodata   & \nodata    & 4197 & 74\\
HIP  25428  &  2013 Nov 20  &  $1.1\pm0.2$  & \nodata   & \nodata    & 1125 & 120\\
HIP  53985  &  2013 Dec 24  &  $1.0\pm0.1$  & 2499  & 75     & 2517 & 74\\
HIP  13402  &  2013 Dec 24  &  $0.9\pm0.1$  & 3378  & 38     & 3395 & 26\\
  HD 78141  &  2013 Dec 24  &  $0.9\pm0.0$  & 1092  & 57     & 920  & 57\\
HIP  44901  &  2013 Dec 25  &  $1.2\pm0.1$  & 3467  & 84     & 3493 & 83\\
HIP  53910  &  2013 Dec 25  &  $1.0\pm0.1$  & 3376  & 63     & 3375 & 61\\
HIP   6061  &  2013 Dec 25  &  $1.1\pm0.2$  & 3353  & 65     & 3383 & 65\\
HIP  10552  &  2013 Dec 26  &  $2.0\pm0.3$  & \nodata   & \nodata    & 1685 & 34\\
HIP  58001  &  2013 Dec 26  &  $2.0\pm0.3$  & 105   & 3      & 105  & 3\\
HIP  62933  &  2013 Dec 26  &  $1.4\pm0.3$  & 3677  & 51     & 3691 & 44\\
HIP  44127  &  2013 Dec 26  &  $1.8\pm0.7$  & \nodata   & \nodata    & 1870 & 28\\
HIP  28360  &  2013 Dec 27  &  $1.4\pm0.1$  & 3018  & 86     & 3015 & 80\\
HIP  44458  &  2013 Dec 27  &   1.6        & 1760  & 89     & 1776 & 103\\
HIP  65327  &  2013 Dec 27  &  $2.1\pm0.4$  & 2328  & 47     & 2296 & 47\\
HIP   8497  &  2013 Dec 27  &  $1.8\pm0.3$  & 3401  & 44     & 3404 & 44\\
HIP  33202  &  2013 Dec 28  &  $1.8\pm0.5$  & 3150  & 60     & \nodata  & \nodata\\
HIP  15457  &  2013 Dec 30  &  $1.6\pm0.2$  & 2722  & 47     & 2713 & 47\\
HIP  66249  &  2013 Dec 30  &  $2.0\pm0.2$  & 2161  & 64     & \nodata  & \nodata\\
HIP  36188  &  2013 Dec 30  &  $1.6\pm0.2$  & 3028  & 54     & 3077 & 57\\
HIP  21547  &  2013 Dec 31  &  $1.3\pm0.2$  & 2928  & 47     & 3017 & 47\\
BD+20 1790  &  2014 Feb 10  &  $1.3\pm0.1$  & 1517  & 54     & \nodata  & \nodata\\
HIP  63076  &  2014 Feb 14  &  $1.0\pm0.1$  & 3276  & 49     & 3272 & 50\\
HIP  72603  &  2014 Feb 14  &  $1.4\pm0.1$  & 2649  & 25     & 2655 & 24\\
HIP  48341  &  2014 Feb 14  &  $1.2\pm0.1$  & 4073  & 50     & 4066 & 42\\
HIP  25486  &  2014 Feb 14  &  $1.2\pm0.1$  & \nodata   & \nodata    & 3330 & 32\\
HIP  80459  &  2014 Mar 12  &  $0.9\pm0.1$  & 2002  & 41     & 1978 & 40\\
HIP  41152  &  2014 Mar 12  &  $1.0\pm0.1$  & 2857  & 51     & 2857 & 50\\
HIP  58001  &  2014 Mar 12  &  $1.4\pm0.3$  & 4078  & 49     & 4080 & 51\\
HIP  56997  &  2014 Mar 13  &   0.9        & 2419  & 169    & 2413 & 170\\
HIP  85829  &  2014 May 08  &  $1.1\pm0.1$  & 2959  & 59     & 2953 & 56\\
HIP  69732  &  2014 May 08  &  $1.3\pm0.2$  & 3994  & 117    & 4008 & 116\\
HIP  62956  &  2014 May 08  &  $1.1\pm0.2$  & 2610  & 38     & 2612 & 38\\
HIP  61960  &  2014 May 09  &  $1.1\pm0.2$  & 2038  & 47     & \nodata  & \nodata\\
HIP  72848  &  2014 May 09  &  $1.0\pm0.1$  & 2643  & 75     & \nodata  & \nodata\\
HIP  91262  &  2014 May 09  &  $0.9\pm0.1$  & 3683  & 158    & \nodata  & \nodata\\
HIP 107556  &  2014 Oct 04  &   0.7        & 3720  & 39     & 3615 & 39\\
HIP  16537  &  2014 Oct 04  &   0.7        & 3856  & 49     & 3822 & 48\\
HIP   8497  &  2014 Oct 04  &   0.6        & 3741  & 41     & \nodata  & \nodata\\
HIP 105199  &  2014 Oct 05  &  $1.5\pm0.2$  & 4008  & 81     & \nodata  & \nodata\\
HIP  19859  &  2014 Oct 05  &  $0.8\pm0.1$  & 2767  & 60     & 2764 & 60\\
BD+48 3686  &  2014 Oct 06  &  $0.8\pm0.1$  & 3981  & 108    & 3980 & 108\\
HIP 105918  &  2014 Nov 06  &  $1.3\pm0.2$  & 2209  & 44     & \nodata  & \nodata\\
HIP   2843  &  2014 Nov 06  &  $1.7\pm0.3$  & 1375  & 45     & \nodata  & \nodata\\
HIP 111278  &  2014 Nov 08  &  $1.1\pm0.2$  & \nodata   & \nodata    & 1782 & 54\\
HIP   8486  &  2014 Nov 08  &  $1.2\pm0.1$  & 781   & 6      & 765  & 10\\
HIP  12828  &  2014 Nov 08  &  $1.1\pm0.2$  & 2508  & 36     & 2509 & 32\\
HIP  36704  &  2014 Dec 11  &   0.8        & 4125  & 153    & 4128 & 153\\
HIP   1473  &  2014 Dec 11  &   1.0        & 1987  & 84     & 1989 & 86\\
HIP  10064  &  2015 Jan 05  &  $0.9\pm0.1$  & 1463  & 146    & 1449 & 146\\
BD+20 1790  &  2015 Jan 05  &  $0.8\pm0.1$  & 3481  & 80     & 3470 & 79\\
HIP  21547  &  2015 Jan 05  &  $0.9\pm0.1$  & 2997  & 34     & \nodata  & \nodata\\
HIP  63125  &  2015 Jan 05  &   0.9        & 801   & 30     & 784  & 33\\
HIP  58876  &  2015 Jan 06  &  $1.2\pm0.1$  & 3065  & 44     & 3052 & 44\\
HIP  20901  &  2015 Jan 06  &  $1.0\pm0.2$  & 4794  & 79     & 4787 & 82\\
HIP  49669  &  2015 Jan 06  &  $0.9\pm0.1$  & 3535  & 61     & 3539 & 58\\
BD+20 1790  &  2015 Jan 06  &  $0.9\pm0.1$  & 3032  & 54     & 3017 & 55\\
HIP  23875  &  2015 Feb 08  &  $0.9\pm0.1$  & 2208  & 121    & 2202 & 121\\
HIP  22845  &  2015 Feb 09  &  $1.0\pm0.2$  & 3533  & 69     & 3595 & 65\\
HIP  41820  &  2015 Feb 09  &  $1.6\pm0.2$  & 1593  & 106    & 1591 & 107\\
HIP  63503  &  2015 Mar 06  &  $1.2\pm0.2$  & 1880  & 33     & \nodata  & \nodata\\
HIP  63503  &  2015 Mar 10  &  $1.1\pm0.2$  & 2456  & 48     & 2457 & 49\\
HIP  74702  &  2015 Mar 10  &  $1.5\pm0.2$  & 3599  & 65     & 3583 & 65\\
HIP  43625  &  2015 Mar 10  &  $1.1\pm0.1$  & 3111  & 64     & 3108 & 76\\
HIP  97165  &  2015 Jun 11  &  $1.2\pm0.2$  & \nodata   & \nodata    & 598  & 10\\
HIP  93805  &  2015 Jun 11  &  $1.4\pm0.1$  & 839   & 29     & 836  & 28\\
HIP 109427  &  2015 Jun 12  &  $0.9\pm0.1$  & 3222  & 45     & 3233 & 44\\
 HD 106591  &  2015 Jun 12  &  $1.0\pm0.1$  & 264   & 9      & 264  & 9\\
HIP 114570  &  2015 Jun 26  &  $0.9\pm0.1$  & \nodata   & \nodata    & 1295 & 29\\
HIP  44897  &  2016 Mar 25  &  $1.2\pm0.2$  & 3248  & 169    & 3241 & 169\\
HIP  71075  &  2016 Mar 26  &  $1.3\pm0.2$  & 2753  & 25     & 2769 & 25\\
HIP  61481  &  2016 Mar 27  &  $1.2\pm0.2$  & 2236  & 47     & 2233 & 47\\
HIP  61481  &  2017 Apr 05  &  $1.1\pm0.2$  & 3121  & 56     & 3115 & 55\\
\enddata 
\tablenotetext{a}{Mean and standard deviation of the seeing as measured by the
DIMM at LBT and recorded in image headers. For some data sets, seeing was
unavailable in headers. For these we report the value written in the nightly
observing log.}
\end{deluxetable}

\section{Data Reduction and Processing} \label{DataProcessing}
\subsection{Basic Image Processing} 
We use a bad-pixel mask to identify pixels with persistent problems and correct
them using the median of the nearest
8 good pixels. LMIRcam images exhibit low-level time-variable offsets in each
  $64\times1024$ pixel readout channel. We remove these offsets by subtracting
the median pixel value from each channel after excluding  all 3$\sigma$
outliers. All images are corrected for distortion using a 2-dimensional
polynomial transformation and the coefficients reported by \citet{Maire2015}.
After distortion correction, image pixels are binned 2x2. Binning the over
sampled frames reduces the influence of any cosmic ray hits during the exposure
and any bad pixels not represented in the mask.  

We remove variable sky-background using nod subtraction. Although the left and
right sides of the telescope are typically nodded together during dual-sided
observations, nods on each side are fundamentally independent. Thus, we
distinguish between the left and right image when performing nod subtraction.
This gives us added flexibility to utilize frames when one side of the
telescope is experiencing issues (e.g., an open AO loop).  

After nod subtraction, each beam (left-up, left-down, right-up, right-down) is
cropped to a $3\arcsec$ radius field of view. Then, all images within a beam are
co-aligned on a subpixel level using a cross-correlation with a median-combined
image as a template. After shifting all images, the cross-correlation process
is repeated with a new median template.  Images with the least correlation with
the second median image are discarded. This effectively filters images with
poor Strehl ratio and open loop data. We typically reject the worst 10\% to
30\% of images depending on the quality of the data and the AO-loop stability.
Exposure times in Table \ref{ObservingLog} report the total retained data after
culling.

\subsection{High-contrast Image Processing}\label{highContrast}
We reduce each nod position (up/down left/right) independently, combining the
beams to produce a master image using a weighted mean as a last step. Keeping
the positions separate produces similarly reduced subsets of our data that
allow us to better distinguish real signals from time-variable speckles.  In
some of our data sets some nod positions exhibit diffraction rings from dust
contamination on optics near the focal plane in LMIRcam. In these cases, our
approach allows us to downweight the nod positions with contamination in order
to maximize our contrast.

For each nod position, we combine the reduced images into groups with less than
$2^{\circ}$ of parallactic angle change. This rotation limit is chosen so that
a source at $3\arcsec$ moves $\lesssim \lambda/D$. These combined images are
then used in our principal component analysis (PCA)-based image processing
algorithm. 

In PCA-based image processing algorithms, the stellar point-spread function
(PSF) and quasi-static speckles in each image are modeled as the projection
onto the subspace spanned by the leading eigenimages of the pixel covariance
matrix. We use all combined images as reference images for the application of
PCA. The signal of a faint, rotating planet is not expected to be represented
in one of the leading eigenimages. We measured our throughput on a typical
double-sided data set and recovered 35\% at $0\farcs5$ and $68\%$ at $2\arcsec$
(21 $\lambda/D$). To put this in context, another commonly used
high-contrast image processing algorithm ---locally optimized combination of
images (LOCI)--- delivers $\sim65\%$ throughput at 35 $\lambda/D$
\citep{Lafreniere2007B}.

Our approach works annulus by annulus in each image. We model annuli of
13 pixels ($\sim3\frac{\lambda}{D}$) wide but subtract only the central 1 pixel
   wide ring for each radius.   For each radius, the number of principal
components we use to model the stellar PSF and speckle distribution is
optimized to maximize the achieved contrast (see Section \ref{optimize}). After
removing the image of the star, all images are derotated using the parallactic
angle and the offset to true north reported in \citet{Maire2015}. We then
median-combined to enhance signal to noise.  

\subsection{Optimizing NPCs} \label{optimize}

For each radius, we determine the optimal number of principal components
(NPCs) to subtract by injecting fake sources and maximizing the measured
signal-to-noise ratio.  

For the artificial source injections, we use unsaturated images of the primary
star taken before and/or after our saturated imaging as a photometric standard
and PSF model. Some of our targets do not have unsaturated frames saved as part
of their data set. For these targets, we use unsaturated images of another star
taken the same night or, in one case, we use the unsaturated image of an
optical ghost that appears 5.76 mag fainter than the primary. We inject
planets that rotate opposite the sky rotation in our frames and then reduce
them using the opposite of the parallactic angle to derotate in the last step
of our high-contrast image processing algorithm. This reduces our
susceptibility to biases from real astrophysical signals in the data
\citep[e.g.,][]{Wahhaj2013}.

We measure the signal-to-noise ratio of each artificial planet as follows. After PCA
subtraction, we smooth the image with a Gaussian with the same width as our PSF
and then take the peak pixel value in the vicinity of our injected source as
the signal level. For a noise estimate, we process the image stack without
injecting an artificial source but using the same NPCs and reverse rotation.
We then smooth the image using the same kernel as before and take the
noise level to be the standard deviation of pixel values within the 13 pixel
wide annulus centered at the radius of the artificial planet.

In addition to optimizing the NPCs necessary at each radius, this process also
automatically produces a $5\sigma$ contrast curve quantifying our photometric
sensitivity as a function of radial separation from the host star. As discussed
below, these contrast curves are improved when we combine nods and telescope
apertures to create a final image for each target.  In Section
\ref{photometricLimits}, we discuss the necessary adjustments to these contrast
curves to account for the small number statistics at small separations.
Since our contrast curves are created with artificial source injection,
the effects of our algorithmic throughput are built in.

\subsection{Combining Beams} 
\begin{figure}
\plotone{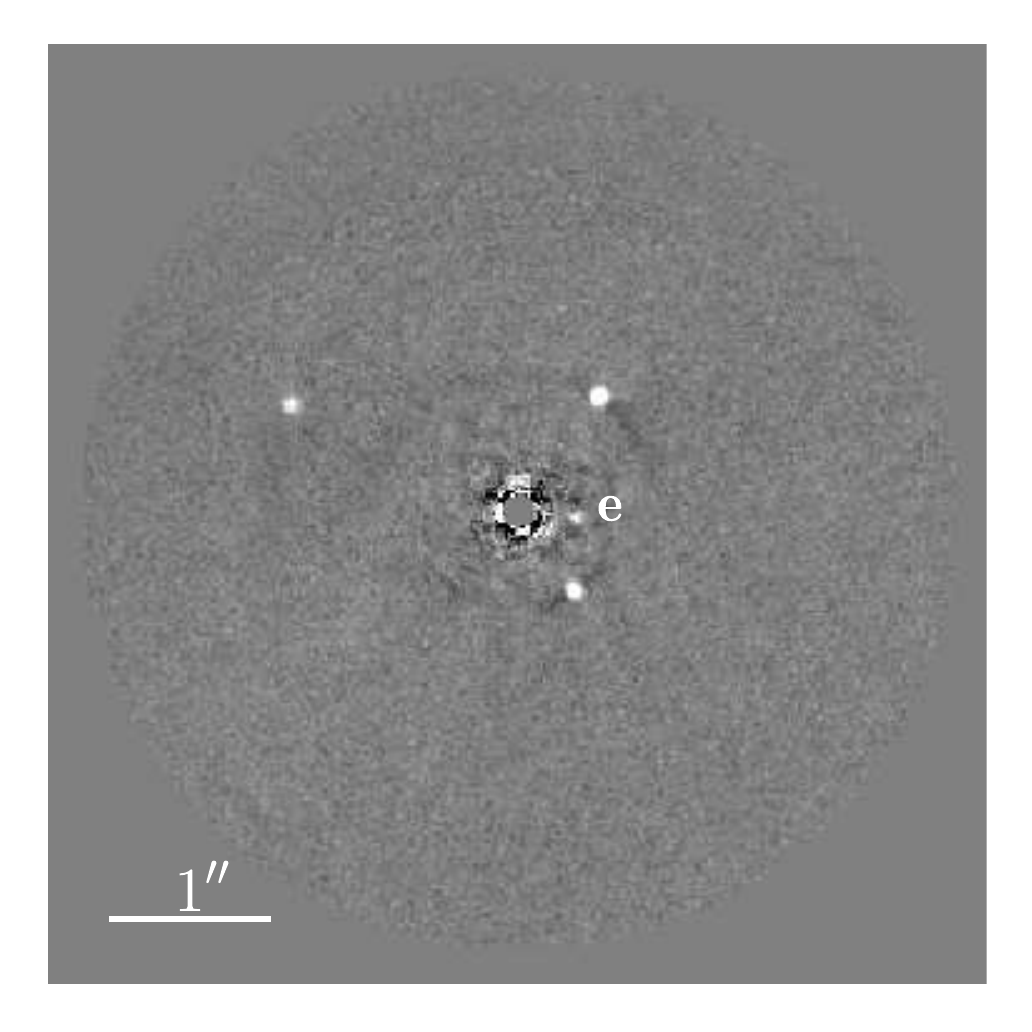}
\caption{Demonstration of the functionality of our LEECH data processing
pipeline on an HR~8799 dataset. We detect all four directly imaged
planets at high significance even though the data were collected using only one
side of the telescope under poor conditions (seeing $\gtrsim1\farcs5$). The
innermost planet ``e" (labeled) is separated by less than $0\farcs4$ and is
detected with a signal-to-noise ratio of 7.5.\label{8799}}
\end{figure}

At this point, we have up to four optimally reduced images and contrast curves
for each dataset (one for each nod position; there will be only two for
single-sided datasets). We use a weighted sum to combine each of the separately
reduced beams (up/down left/right) to make a final combined image and contrast
curve.  To find the best weights, for each radius, we inject a fake planet of
the same magnitude into each beam at the same position. We then reduce all
beams using the previously determined optimal NPCs. Finally, we grid-search for
the weights to optimize the signal-to-noise ratio of the resulting summed
image.  Our final contrast curves for each target are constructed so that the
combined contrast is scaled to reflect the increase in the signal-to-noise
ratio after the weighted sum: 
\begin{equation}
\Delta m(r)_{\mathrm{comb}} = {\Delta}m_{i}(r) + 2.5\log_{10} \left( \frac{\mathrm{SNR}_{\mathrm{comb}}}{5} \right), 
\end{equation} 
where ${\Delta}m_{i}(r)$ is the initial contrast that was the deepest measured
among the beams, $\mathrm{SNR}_{\mathrm{comb}}$ is the signal-to-noise ratio that
resulted after the weighted combination, and ${\Delta}m(r)_{\mathrm{comb}}$ is
our final contrast at radius $r$.  In Figure \ref{8799}, we show a fully
reduced image of the HR~8799 system to demonstrate the performance of our
pipeline. The HR~8799 dataset was taken under poor conditions (seeing
$\gtrsim1\farcs5$) on 2013 October 22 using only the right side of the LBT. The
image includes 44 minutes of exposure time and tracks $80^{\circ}$ of
parallactic angle change. We detect the innermost planet, HR~8799~e ---
separated by $\lesssim0\farcs4$ --- with a signal-to-noise ratio of 7.5.

\section{Results} \label{Results}
\subsection{Search for Planetary Companions}

To search all of our final reduced images for new companions, we use both
visual inspection and automated signal-to-noise maps. Our by-eye approach uses
the individually reduced beam images as well as the final combined image for
each target. This allows us to quickly discriminate between real objects and
bright speckles because astrophysical sources should appear at the same
position in each beam, albeit at a lower signal-to-noise ratio.

We automatically constructed signal-to-noise maps for each target by smoothing
the final combined image to estimate signal strength on the scale of our PSF,
and smoothing reverse-rotated final combined images to estimate the noise level
in each annulus. We flagged and inspected each data set with at least
1 pixel above a signal-to-noise ratio of 4. These maps did not reveal any
  signals not previously flagged by our visual inspection. We did not discover
any new exoplanets orbiting any of our targeted stars.

In the following subsections, we describe our findings in detail.

\subsubsection{HIP 21547 (51 Eri)}

We observed HIP 21547 in 2013 December and in 2015 January. The 2013
dataset was not as sensitive as the 2015 dataset because they were taken during
worse conditions ($1\farcs3$ seeing in 2013 compared to $0\farcs9$ in 2015).
Our 2015 data were collected with good seeing, include 75 minutes of exposure
time, and $34^{\circ}$ of rotation. The target, also known as 51 Eri, has
a directly imaged planet at 0\farcs45 separation with an $L^{\prime}$ contrast of
11.58 \citep{Macintosh2015, Rajan2017}. Ten percent of our contrast curves
attain at least this depth at 0\farcs45.

We recovered the planet reported by \citet{Macintosh2015} at only 2.4$\sigma$
in our 2015 dataset. The low significance of our detection is likely the result
of multiple factors. For one, temporary trouble with the right-side AO
system meant this target was observed in single-sided mode, even though
most targets were observed in double-sided mode on the same night.  Second, the
observing log reports variable thin clouds which likely affected the
thermal-infrared transmission and background flux level. Lastly, we obtained
only $34^{\circ}$ of sky rotation, which corresponds to less than 3($\lambda/D$)
motion at the separation of the planet and is inadequate for obtaining the
deepest contrasts.  Analysis of our observing metadata for the survey revealed
that our best contrast interior to $0\farcs7$ occurs when we obtain
$\gtrsim70^{\circ}$ of rotation.

\subsubsection{New Component in the HIP 97165 ($\delta$~Cyg) System} 

\begin{figure}
\plotone{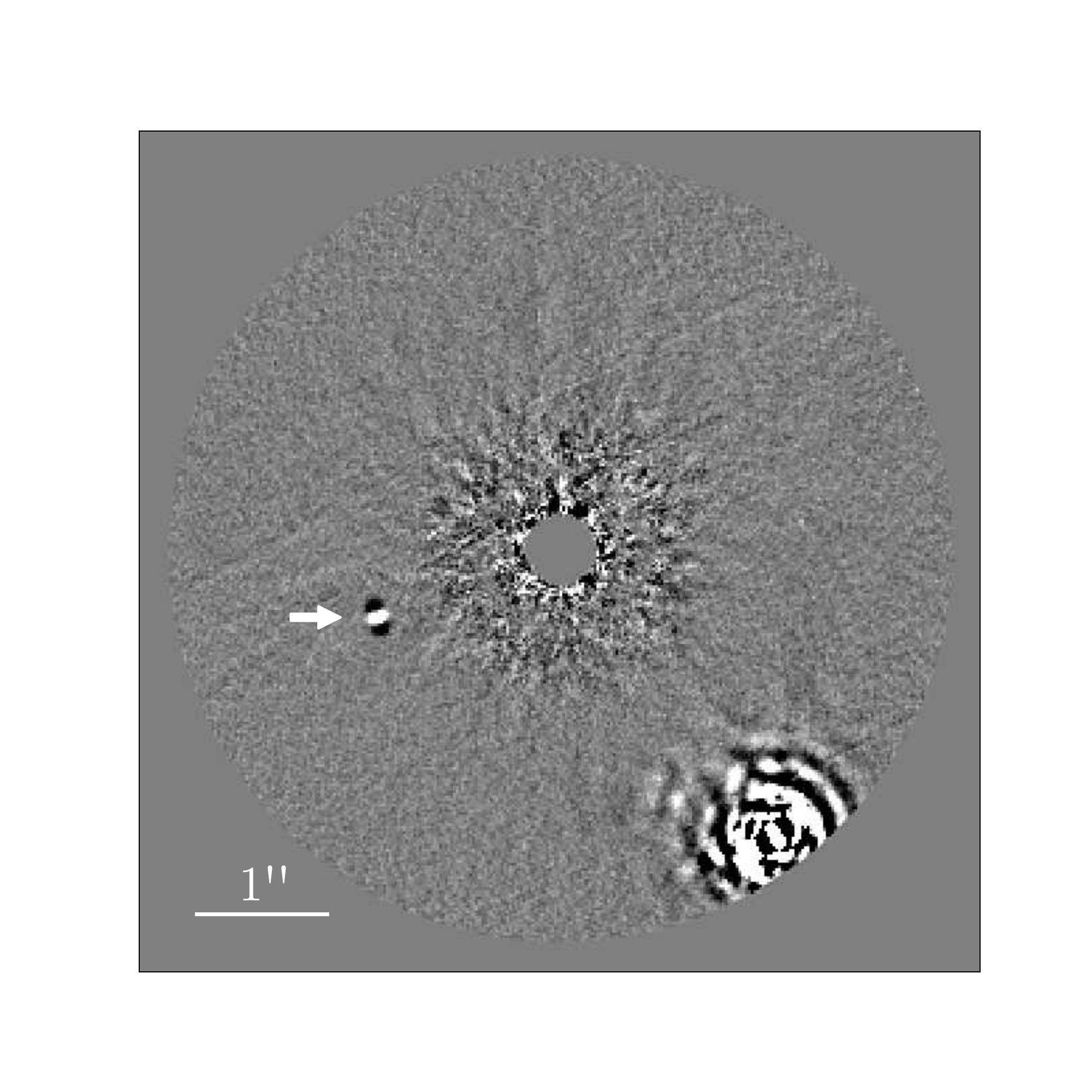}
\caption{A newly discovered component in the $\delta$~Cyg (HIP~97165) system
(indicated with an arrow). North is up and east is left. The object near the
southwest edge of our field of view is $\delta$~Cyg~B.\label{deltaCyg}}
\end{figure}

We discovered a bright source in our $L^{\prime}$ imaging data at a position
angle of $110\degr\pm1\fdg5$ separated by $1\farcs49\pm0\farcs04$ from the
primary in the $\delta$ Cyg system, shown in Figure \ref{deltaCyg}. We observed
the object in additional infrared filters to characterize its color.  We
compared our photometry of this newly discovered object (listed in Table
\ref{deltaCygTable}) to the \citet{Baraffe2015} evolutionary models to estimate
its mass, $\sim0.13$\---$0.2~M_{\odot}$. The $L^{\prime}$ photometry
of the newly discovered star was measured by injecting a scaled negative image
of the unsaturated primary into each of the individual images before performing
the PCA analysis.  For the $H$, $Ks$, and $Ls$ photometry, we measured the
brightness of the object using the secondary as a PSF model and PSF fitting after
subtracting a 180$^{\circ}$ rotated version of the image to remove the
rotationally symmetric portion of the primary star PSF. All photometric errors
are estimated by repeating our measurements several times with a random
selection of the data frames \citep[i.e., bootstrapping; see][]{Press2002}.

While the mass estimates from each band overlap, our photometry suggests
a bluer color than predicted by the models. However, systematic differences in
the data collection and analysis between the $H$, $Ks$, and $Ls$ band images
and the $L^{\prime}$ image complicate our ability to interpret the color.

The $\delta$~Cyg system has high proper motion, $pm_{\mathrm{R.A.}}=38\pm1.4$~mas
yr$^{-1}$, $pm_{\mathrm{Decl.}}=52\pm1.3$~mas yr$^{-1}$ \citep{GAIA2018}, but we
were not able to confirm common proper motion of the companion with our data.
We checked for a similarly bright object in the vicinity of the star in the
2MASS Survey Atlas image taken in 1998. From 1998 to 2015, $\sim1\arcsec$ of
motion is expected. However, the 2MASS images are saturated out to a radius of
$\sim2\farcs5$, making this test impossible. We also checked observatory
archives for other high-contrast imagers, looking for previous observations of
$\delta$ Cyg. We found none.

We calculate the probability of a chance alignment using the sky density of
similarly bright sources in the neighborhood of $\delta$~Cyg from the 2MASS
Point Source Catalog \citep{Cutri2003,Skrutskie2006}. For
a $10^{\circ}\times10^{\circ}$ patch of sky centered on $\delta$~Cyg the
average number of sources with Ks mag less than 12 in a square degree is 631.
The implied number of sources within a $1\farcs5$ radius circle is
$3.4\times10^{-4}$.  

Assuming all our targets sample a similarly populated portion of the sky, then
we expect to have a $\sim3\%$ chance of detecting an $m_{Ks}=12$ or brighter
source within $1\farcs5$ of one of our 98 stars. However, some of our stars are
located in more sparsely populated regions, and some of our stars are located in
more densely populated regions of the sky so our derived number, $\sim3\%$,
should be interpreted with caution.  Additional observations of this object are
warranted to confidently determine its nature.

\begin{deluxetable}{ll}
\tabletypesize{\footnotesize}
\tablecolumns{2}
\tablewidth{0pt}
\tablecaption{New $\delta$ Cyg Companion Photometry\label{deltaCygTable}}
\tablehead{
    \colhead{filter}&
    \colhead{Abs. mag}}
\startdata
$L^{\prime}$ & $8.58\pm0.06$ \\
$Ls$ & $8.63\pm0.12$ \\
$Ks$ & $8.43\pm0.05$ \\
$H$  & $8.38\pm0.05$ \\
\enddata
\end{deluxetable}

\subsubsection{Confirmed Background Objects} 

\begin{figure*}
\includegraphics[width=\textwidth]{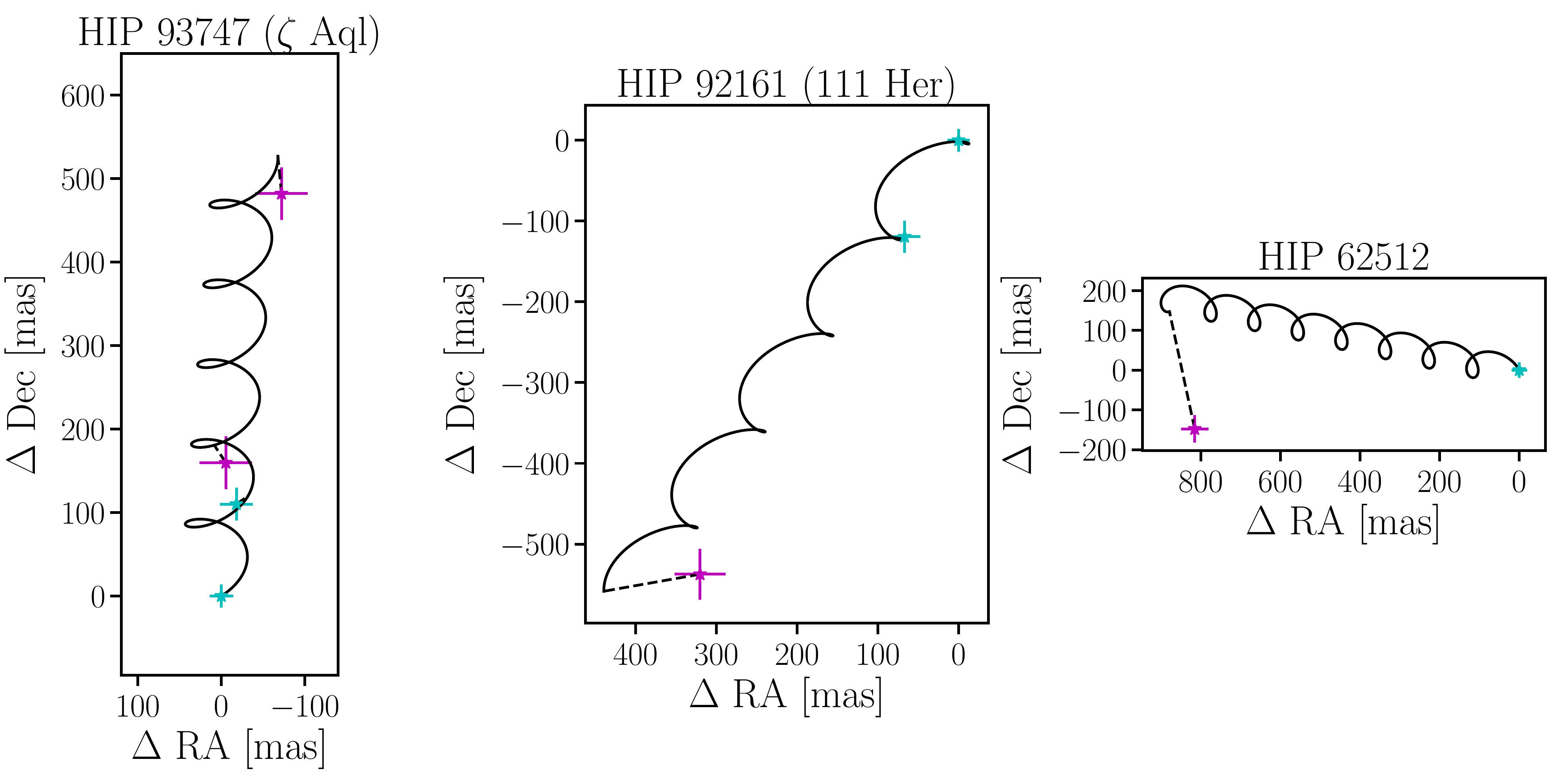}
\caption{Relative astrometry of faint sources in the vicinity of three of our
target stars. Cyan points are from archival NIRC2 data and magenta points are
from LMIRcam.  Solid curves show the expected motion of stationary background
objects with respect to the target star including both proper motion and annual
parallax. Dashed lines connect measured astrometric points to the corresponding
position of the background track at the time observed. The LEECH point for
HIP~62512 is inconsistent with the background track and with Keplerian motion
from a bound orbit. We attribute its offset to low signal-to-noise in the NIRC2
data and working close to the edge of the field of view in the LEECH data.
\label{background}} 
\end{figure*}

Our survey is less susceptible to ambiguous background sources than other
surveys performed at shorter wavelengths because the much brighter
sky background at $L^{\prime}$ limits the volume of the galaxy we probe with
each dataset.  Yet, we find background objects in the vicinity of three of our
targets, HIP~93747 and HIP~92161, which are both at low galactic latitude, and
HIP~62512.  We compared the LEECH-measured positions of each object to archival
Keck NIRC2 data that we reduced to confirm their background nature. Figure
\ref{background} shows the relative astrometry for each object compared to the
motion expected of a distant background source. We use conservative astrometric
error bars based on the analysis of \citet{Bowler2015} for NIRC2 and
\citet{Maire2015} for LMIRcam. Our LMIRcam astrometry includes an additional
error of 1 pixel to account for challenges in centroiding the heavily saturated
PSFs of the primary stars. The astrometry of the source in the vicinity of
HIP~62512 seems inconsistent with the background track presented in Figure
\ref{background}.  The motion of the HIP~62512 source is inconsistent with
a bound Keplerian orbit because the minimum average velocity implied by the
apparent projected motion from 2005 to 2013 exceeds the maximum escape velocity
calculated by assuming the true separation is equal to the projected separation
by a factor $\sim3$. Thus, we attribute the discrepancy with the background
track for this source to systematic errors because the source was imaged at the
edge of our field of view in the LEECH data and is also affected by
a diffraction spike in the NIRC2 data. Additional observations of this system
are warranted to definitively confirm its nature.

\subsection{Survey Sensitivity} 

\subsubsection{Photometric Limits}\label{photometricLimits} 

We present two versions of our contrast curves: (1) a ``classical" set,
which do not correct for small number statistics, and (2) a ``modern" set that
does include a correction for small number statistics \citep{Mawet2014}. Our
``modern" contrast curves indicate a constant number of expected false
positives as a function of separation \citep[][]{Jensen-Clem2018} and ensure
95\% completeness. Our varying threshold allows a total of 0.01 false
detections within $3\arcsec$ from our targets. The ``classical" contrast curves
(described in Section \ref{DataProcessing}) are presented and analyzed for
comparison to previous studies.  We describe how we adjust these to produce our
modern contrast curves in Appendix \ref{Appendix}.  On average, our more
carefully constructed  modern contrast curves are 0.28 magnitudes less
sensitive than our classical curves, though this varies with separation (see
Appendix \ref{Appendix}).

In the left panel of Figure \ref{contrastcurves}, we show the best and median
contrast curves from our survey and compare them to the best and median
contrast curves from the \citet{Rameau2013} survey, which was also conducted at
$L^{\prime}$. Comparing like to like, our median classical $5\sigma$ contrast
is $\sim1$~mag deeper across the whole range of angular separations probed
compared to \citet{Rameau2013}.  Our best contrast curve is similar to the best
curve reported by \citet{Rameau2013}.  In the right panel of Figure
\ref{contrastcurves}, we show median contrast versus projected physical
separation assuming a distance equal to the respective survey median target
distances. In this case, the performance difference is more pronounced,
revealing more than 3~mag of increased contrast interior to 12~au
projected separation.

\subsubsection{Sensitivity in the Mass\---Semi-Major Axis Plane} \label{sensitivitySection}
While median contrast curves can provide a good metric for comparing
photometric and AO performance, it does not indicate sensitivity
to planets. This is because the stellar age and magnitude for each target must
be taken into account to convert contrast to limiting magnitude and limiting
magnitude to an upper limit on planetary mass with the assistance of an
evolutionary model. 

Evolutionary models predict how planets of different masses cool and fade over
time. We choose to use three different models to derive three separate
estimates of our sensitivity to gas-giant exoplanets.  These models are DUSTY
\citep{Chabrier2000}, COND \citep{Baraffe2003}, and the models presented by
\citet{Fortney2008}, which we will refer to as F08.  Each of the three
evolutionary models is built on a different set of extreme assumptions for
formation, evolution, and the atmospheric appearance of substellar objects.
None of the models seem to be precise fits to the observed directly imaged
planet population, but taken together, they bracket the data in color-magnitude
space.

The DUSTY and COND models are both `hot-start' in the sense that they make use
of arbitrarily large adiabatic spheres that undergo homologous collapse as
their initial condition. These models produce bright young objects
\citep{Baraffe2002, Marley2007}. More physically motivated models for the early
evolution of gas-giant luminosities are based on the core-accretion scenario
\citep[e.g.,][]{Marley2007,Mordasini2013}. These models can produce a variety
of initial post-accretion luminosities depending on the radiative efficiency of
the accretion shock while envelope material is being built up
\citep[e.g.,][]{Mordasini2017}. The core-accretion formation prescription
included in the F08 models \citep[that of ][]{Hubickyj2005} appears to
produce very-low luminosity young planets \citep{Berardo2017}, so we take our
F08 sensitivity to be conservative. 

The DUSTY and COND models represent atmospheric extremes with respect to dust
and cloud opacity. DUSTY atmospheres exhibit maximal dust opacity, retaining in
the photosphere all the dust and condensates that form. COND models assume no
photospheric dust opacity, but assume that dust forms and immediately
precipitates below the photosphere (taking its constituent molecular species
with it). The F08 cold-start models are cloud-free.

Model fits to photometric and spectroscopic measurements of the HR~8799 planets
and other young low-gravity gas giants reveal that planets can loft clouds even
at effective temperatures where higher-gravity brown dwarfs appear mostly cloud
free \citep[down to $\sim900$ K; e.g.,][]{Currie2011, Barman2011, Skemer2014, Ingraham2014,
Bonnefoy2016}.  Unfortunately, the cloud-containing DUSTY models do not extend
to the intermediate ages of all of our targets. The DUSTY models truncate
earlier for lower-mass objects (e.g., 40 Myr for $2~M_{\mathrm{Jup}}$, 300 Myr
for $6~M_{\mathrm{Jup}}$, 1000 Myr for $10~M_{\mathrm{Jup}}$). In order to
better capture the dusty faint appendix to the L-dwarf sequence seen for
low-gravity atmospheres \citep[e.g.,][]{Liu2016}, we extrapolated the DUSTY
models by enforcing that they remain parallel to the COND models but with the
same offset as measured for the last age at which both models include
predictions. We caution that for the coolest objects, this extrapolation becomes
a very poor approximation to the observed flux. The atmosphere of 51~Eri~b
($T_{\mathrm{eff}}\sim700$~K) appears to be significantly less cloudy than
hotter low gravity objects, so the cloudy extension to low-gravity does not
continue to temperatures below $\sim700~K$ \citep[although some cloudiness is
necessary to account for the observed $L^{\prime}$ flux of 51~Eri, these are
likely not the silicate clouds modeled in the DUSTY grid;][]{Macintosh2015,
Rajan2017, Samland2017}. Thus, our extrapolation of the DUSTY models is highly
suspect at lower masses. The accuracy of all the evolutionary models we use is
limited by the incomplete treatment of the relevant physics---including the
initial entropy, the behavior and appearance of clouds, and the treatment of
atmospheric dynamics and disequilibrium chemistry---and all of these issues tend
to become more severe at lower masses and effective temperatures. Concerning
just atmospheric cloudiness, DUSTY models are probably more appropriate than
the COND models for hotter planets, while the COND models are more appropriate
for cooler planets. 

Gas-giant exoplanets are observed to have high mean metal content
\citep{Thorngren2016}, and this manifests, at least partially, in
metal-enhanced atmospheres \citep{Wong2004,Skemer2016,Samland2017}.  The
evolutionary models we use make different assumptions about the composition of
gas-giant planet atmospheres. The \citet{Fortney2008} models are five times
solar metallicity, while both the DUSTY and COND models are solar abundance.
All three models assume equilibrium atmospheric chemistry, which is not well
supported by observations \citep[e.g.,][]{Hinz2010,Barman2011,Skemer2012}.
\citet{Fortney2008} experiment with disequilibrium chemistry in their models,
which show a dramatic increase in the emission at $L^{\prime}$ for atmospheres
with effective temperature greater than 500 K due to the reduction of methane
opacity when mixing within the photosphere delivers carbon monoxide at a rate
faster than chemical reaction timescales \citep{Hubeny2007}.  This effect is
not included in the F08 evolutionary models that we make use of here.

\begin{figure*}
\includegraphics[width=\textwidth]{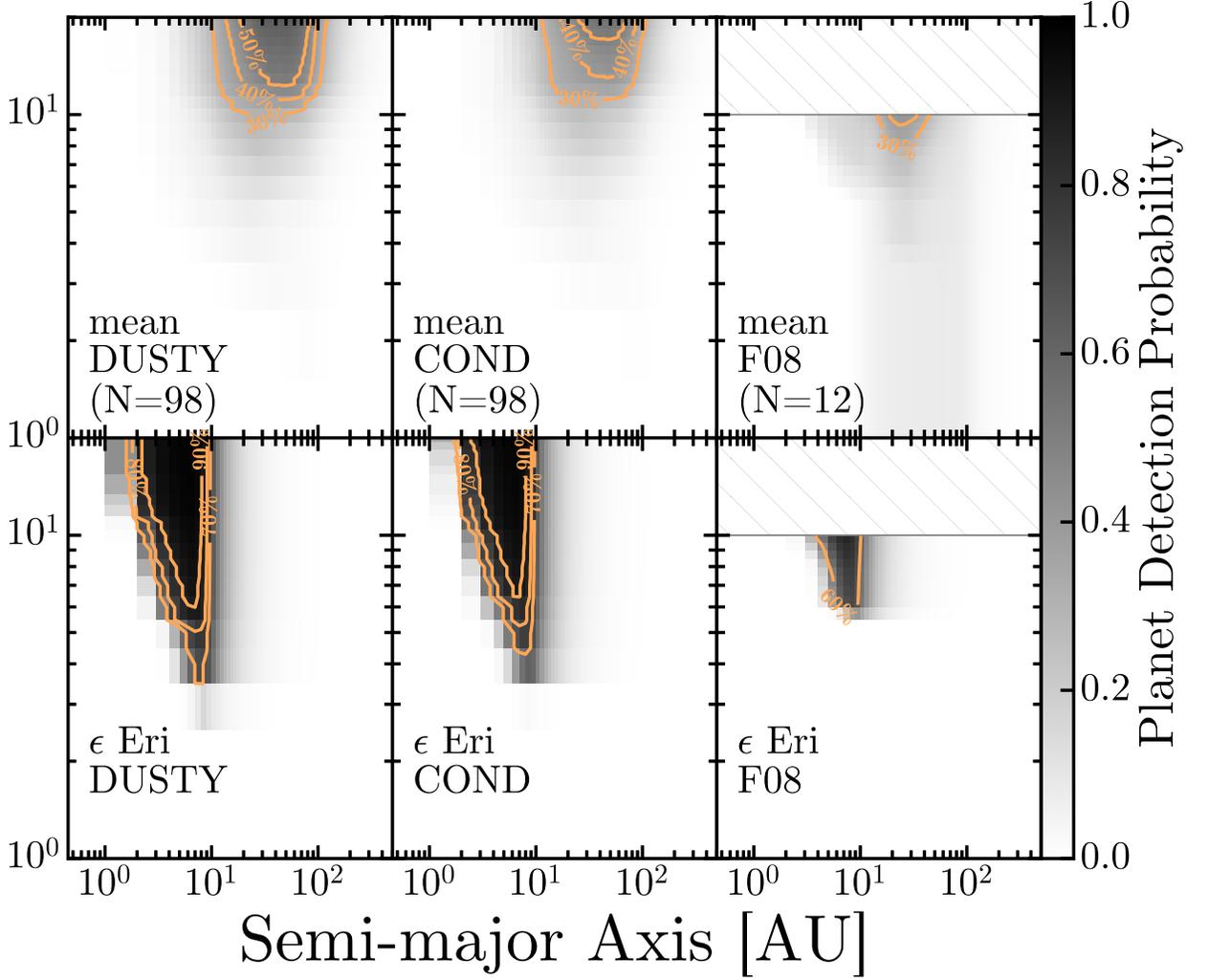}
\caption{Sensitivity maps showing planet detection probability as quantified
using MESS (grayscale and contours). The evolutionary model used is indicated
in each panel. The F08 models do not extend above $10~M_{\mathrm{Jup}}$. Top
row: the mean sensitivity maps averaged over all targets. The number of targets
contributing to the mean is indicated.  Since we were only sensitive to
cold-start planets around 12 stars, the number of targets used to create the
mean F08 map is 12. Bottom Row: Sensitivity maps for $\epsilon$~Eri.}
\label{ModelSensitivities} 
\end{figure*}

For the evolutionary model ($j$) and target star ($i$), we produce a sensitivity
map, $s_{ij}(a, m)$, that quantifies our ability to detect planets as
a function of orbital semi-major axis, $a$, and planet mass $m$.  To produce
these maps, we use the MESS code \citep{Bonavita2012}.  Briefly, MESS performs
Monte Carlo injections and takes $s_i(a, m)$ to be the fraction of planets of
mass $m$ on randomly oriented Keplerian orbits with semi-major axis $a$ that
could be detected around star $i$ with our data. MESS uses evolutionary models
to convert $m$ to an apparent $L^{\prime}$ brightness given the system age and
distance and then registers a detection if the $L^{\prime}$ brightness is
detectable at the projected separation given our contrast curve. For this
purpose, we generally used our modern contrast curves. We used classical
contrast curves only for the purposes of comparing to previous studies.

MESS treats binary stars differently than single stars, respecting dynamical
constraints on the position of planets following \citet{Holman1999}.
Twenty-six binaries are indicated in Table \ref{TargetSummary} and binary
system parameters are summarized in Table \ref{BinaryTable}. We list the
critical semi-major axis for circumstellar and circumbinary planets in each
system. Circumstellar planets are only allowed interior to $a_{crit}^{cs}$ and
circumbinary planets are only allowed exterior to $a_{crit}^{cb}$. The binary
mass ratio, orbital semi-major axis, and eccentricity are all required to derive
$a_{crit}$. For this purpose, we glean orbital parameters from the Ninth
Catalog of Spectroscopic Binary Orbits\footnote{http://sb9.astro.ulb.ac.be/}
and the Washington Double Star Catalog\footnote{http://ad.usno.navy.mil/wds}.
When information on the flux of the secondary star is available, we derive the
secondary mass in the same way as for the primaries. When there is no
information about the mass of the secondary, we assume a mass ratio of 1. This
is the most conservative assumption because it maximizes the excluded parameter
space. When binary semi-major axis information is missing, we derive the value
using the system period and component masses. When the binary eccentricity is
unavailable, we assume a value of 0.5, following \citet{Bonavita2016}.

In total, 19 out of 26 binaries in the LEECH survey have some portion of the
1 to 500~au range of orbital semi-major axes explored by MESS excluded. The
remaining systems consist of such close binaries that the critical radius is
within 1~au.  Because dynamics excludes the existence of planets at some
positions, we emphasized targeting binaries during intervals of poor observing
conditions to minimize the impact of relatively poor AO performance.

\begin{deluxetable*}{llllllllc}
\tabletypesize{\footnotesize}
\tablecolumns{9}
\tablewidth{0pt}
\tablecaption{Summary of Binary System Parameters\label{BinaryTable}}
\tablehead{
    \colhead{Name}& 
    \colhead{M1}&
    \colhead{M2}&
    \colhead{M2/M1\tablenotemark{a}}  &
    \colhead{Sep.}&
    \colhead{ecc.}&
    \colhead{$a_{crit}^{cs}$\tablenotemark{b}}&
    \colhead{$a_{crit}^{cb}$\tablenotemark{c}}&
    \colhead{Note}\\
    \colhead{}&
    \colhead{($M_{\odot}$)}&
    \colhead{($M_{\odot}$)}&
    \colhead{}&
    \colhead{(\arcsec)}&
    \colhead{}&
    \colhead{(au)}&
    \colhead{(au)}&
    }
\startdata
HIP8486 \tablenotemark{d} & 1.05&  \nodata & 1    &  0.53    &  0.45   &  \nodata &    44.5   &                       \\
HIP8832 \tablenotemark{d} & 2.17& 2.7      & 1    &  8       &  \nodata&  109.7   &   972.0   &                       \\
HIP8903 \tablenotemark{d} & 2.08& 1.2      & 0.57 &  \nodata &  \nodata&  \nodata &    1.6    &                       \\
HIP10064\tablenotemark{d} & 2.39&  \nodata & 1    &  0.01    &  0.44   &  \nodata &    1.1    &                       \\
HIP10552\tablenotemark{d} & 0.85& 0.75     & 0.88 &  0.35    &  0.59   &  \nodata &    48.0   &                       \\
HIP12828\tablenotemark{d} & 1.73&  \nodata & 1    &  \nodata &  \nodata&  \nodata &    4.5    &   1202.2~day period    \\
HIP14576                          & 3.48& 1.70     & 0.48 &  \nodata &  0.26   &  \nodata &    8.7    &  $a_{crit}$ from \citet{Bonavita2016} \\
HIP25486\tablenotemark{d} & 1.06& 0.76     & 0.71 &  \nodata &  \nodata&  \nodata &    10.0   & $a_{crit}$ from \citet{Bonavita2016} \\
HIP27913\tablenotemark{d} & 1.1 &  \nodata & 1    &  0.09    &  0.45   &  \nodata &    11.0   &                       \\
HIP28360                          & 2.79&  \nodata & 1    &  0.003   &  0.0    &  \nodata &    0.2    &                       \\
HIP41820\tablenotemark{d} & 1.00& 0.75     & 0.75 &  0.41    &  0.71   &  \nodata &    45.9   &                       \\
HIP44458\tablenotemark{d} & 1.11& 0.56     & 0.50 &  1.72    &  \nodata&  17.4    &   126.3   &                       \\
HIP49669                          & 3.4 & 0.3      & 0.09 &  \nodata &  0      &  \nodata &    0.7    & $a_{crit}$ from \citet{Bonavita2016} \\
HIP58001\tablenotemark{d} & 2.62& 1.25     & 0.48 &  0.46    &  0.3    &  \nodata &    52.8   &                       \\
HIP62512\tablenotemark{d} & 1.3 &  \nodata & 1    &  0.04    &  0.14   &  \nodata &    3.3    &                       \\
HIP63503\tablenotemark{d} & 1.55& 0.90     & 0.58 &  1.21    &  0.39  &  5.9     &   106.9   &                       \\
HIP65378\tablenotemark{d} & 2.67& 0.5      & 0.19 &  0.78    &  0.6    &  \nodata &    82.8   &                       \\
HIP72603\tablenotemark{d} & 1.4 & 0.9      & 0.64 &  0.4     &  \nodata&  \nodata &    23.0   &                       \\
HIP72659\tablenotemark{d} & 1   & 0.7      & 0.7  &  4.94    &  0.51   &  \nodata &    126.3  &                       \\
HIP72848                          & 0.9 &  \nodata & 1    &  0.02    &  0.51   &  \nodata &    0.7    &                       \\
HIP76267                          & 2.58& 0.92     & 0.35 &  \nodata &  0.37   &  \nodata &    0.7    & $a_{crit}$ from \citet{Bonavita2016}  \\
HIP83207                          & 2.91&  \nodata & 1    &  \nodata &  \nodata&  \nodata &    0.1    &   4.02~day period      \\
HIP85829                          & 1.62&  \nodata & 1    &  \nodata &  \nodata&  \nodata &    0.48   &   38.1~day period      \\
HIP86032\tablenotemark{d} & 2.10& 1.16     & 0.55 &  0.43    &  0.92   &  \nodata &    27.73  &                       \\
HIP97165\tablenotemark{d} & 3.14& 1.55     & 0.49 &  3.41    &  0.52   &  23.4    &    678.8  &                       \\
HIP107556                         & 1.5 & 0.56     & 0.37 &  \nodata &  0.1    &  \nodata &    0.06   & $a_{crit}$ from \citet{Bonavita2016} \\
\enddata 
\tablenotetext{a}{We conservatively assume a mass ratio of 1 when there is no constraint on the secondary mass. This maximizes the excluded parameter space}
\tablenotetext{b}{The circumstellar critical radius. Planets are dynamically excluded on orbits with larger semi-major axes, following \citet{Holman1999}.}
\tablenotetext{c}{The circumbinary critical radius. Planets are dynamically excluded on orbits with smaller semi-major axes, following \citet{Holman1999}.} 
\tablenotetext{d}{These targets include some restricted parameter space in the LEECH sensitivity maps.} 
\end{deluxetable*}

In Figure \ref{ModelSensitivities}, we show examples of the MESS output for each
evolutionary model using our modern contrast curves. Both the mean sensitivity
map averaged over all targets and an example of a sensitivity map for
a specific target are shown. The number of targets used to make the average
maps is indicated in each panel. For the F08 models, we were only sensitive to
planets around 12 of our stars, so we only average those targets. The hashed
region of the F08 panels in Figure \ref{ModelSensitivities} indicates the
portion of parameter space that we did not explore with MESS.

As expected, we are more sensitive to hot-start planets like those modeled in
the DUSTY and COND grids. Of the two, we are most sensitive to DUSTY planets,
since these objects emit a greater fraction of their luminosity in the
$L^{\prime}$-band because cloud opacity in the photosphere pushes the emission
spectrum to longer wavelengths.

The peak of our mean sensitivity is centered between $\sim30$ and $\sim50$~au,
but as suggested in Figure \ref{ModelSensitivities}, the range of semi-major
axes probed is different for each star and depends on the distance.
Indeed, our mean maps indicate significant ($\gtrsim10\%$) sensitivity
to companions $\lesssim20~M_{\mathrm{Jup}}$ extending in to $\sim5$~au.

\section{Statistical Analysis} \label{Analysis} 

\begin{figure*}
\includegraphics[width=\textwidth]{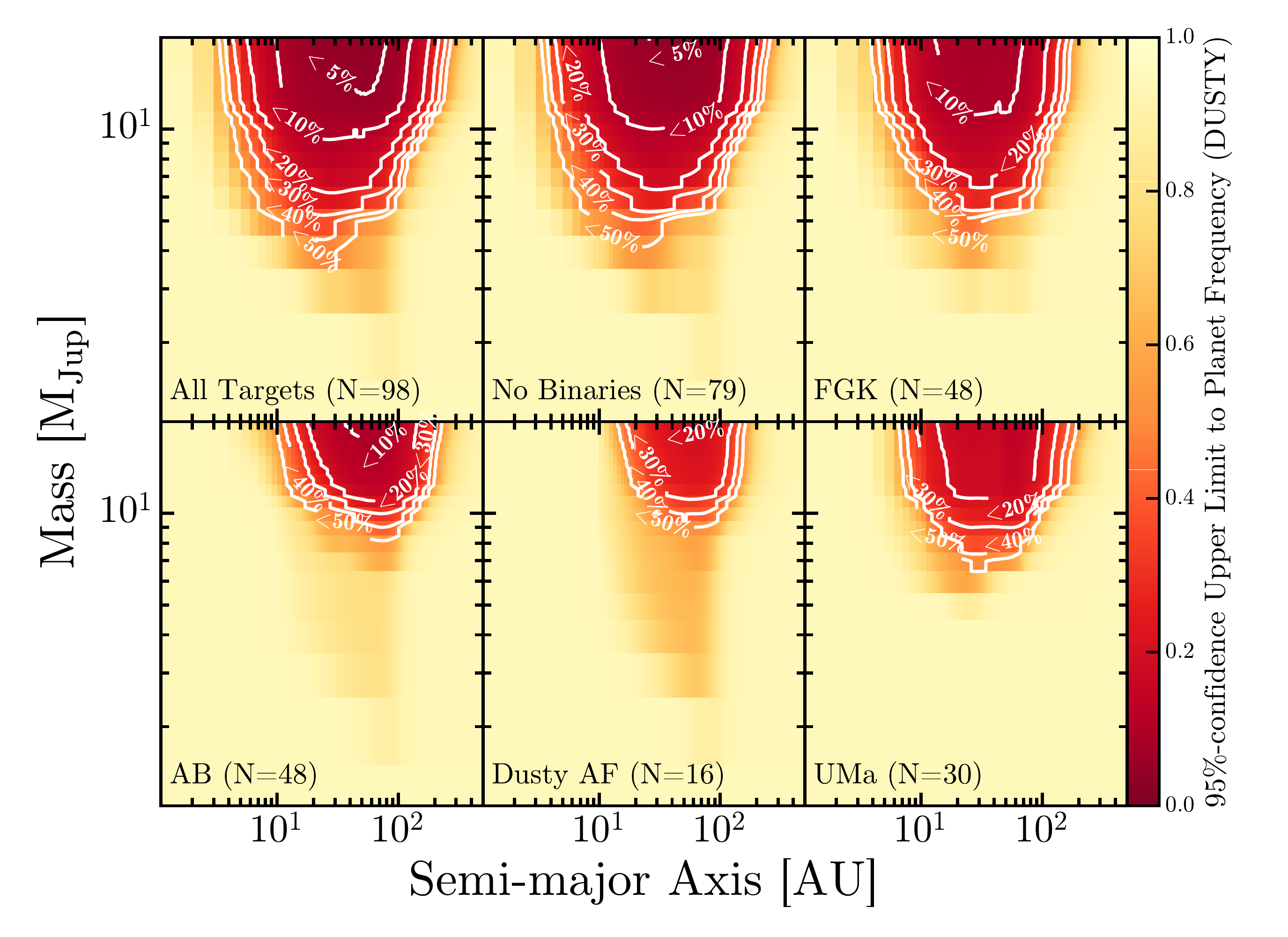}
\caption{The LEECH 95\% confidence upper limits to the planet occurrence
frequency over a fine grid in the mass\----semimajor axis plane calculated
using the DUSTY evolutionary models extrapolated as described in the text (red
colorscale and white contours). We subdivide our sample into groups as
described in the text.\label{dustySubgroups}} 
\end{figure*}

\begin{figure*}
\includegraphics[width=\textwidth]{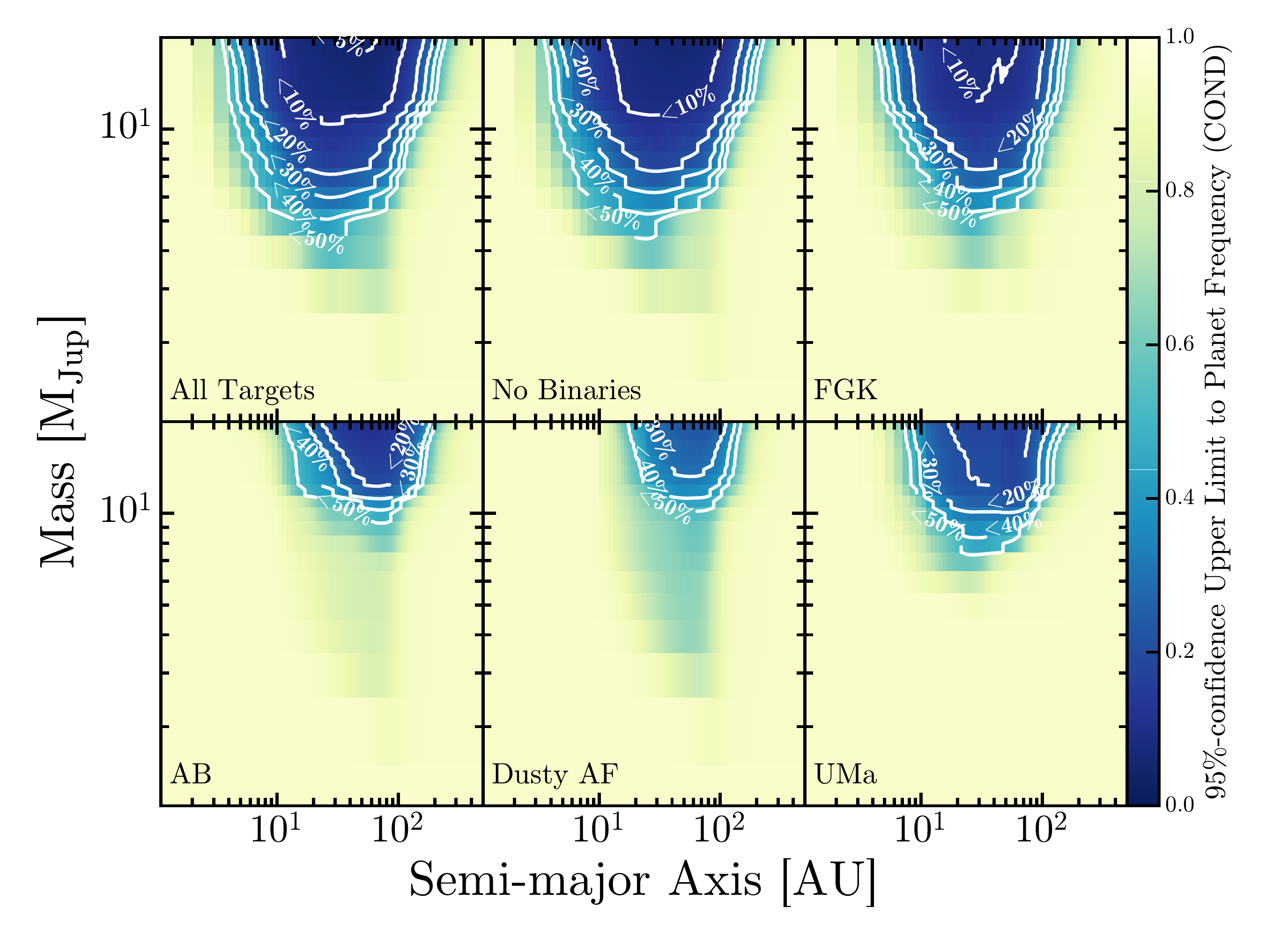} 
\caption{The LEECH 95\% confidence upper limits to the planet occurrence
frequency over a fine grid in the mass\----semimajor axis plane calculated
using the COND evolutionary models (blue colorscale and white contours). We
subdivide our sample into groups as described in the
text.\label{condSubgroups}} 
\end{figure*}

We use LEECH sensitivity maps to derive statistical constraints on the
gas-giant planet occurrence frequency at wide separations.

We did not discover any new planets with the LEECH survey, but as we will show,
we can use our unique sensitivity to place the best-yet limits on cold-start
planets interior to 20~au around FGK stars. We also show improved sensitivity
to hot-start planets interior to $\sim$10-20~au compared to many previous
surveys.

Following the statistical formalism of \citet{Lafreniere2007}, for zero new
discoveries $d$, we adopt the likelihood function
\begin{equation}\label{likelihood}
L(d=0|f) = \prod_{i}^{N} (1-fs_{i}),
\end{equation}
which takes the probability of detecting a planet around the $i$th star to be
the planet occurrence frequency, $f$, times our measurement sensitivity
determined for that target $s_{i}$. With the likelihood function
in Equation \ref{likelihood} we can calculate a posterior distribution using
Bayes theorem,
\begin{equation}\label{posterior}
p(f|d=0) = \frac{L(d=0|f) p(f)}{\int_{0}^{1}L(d=0|f)p(f)df},
\end{equation}
where $p(f)$ is the prior probability distribution for $f$. The 
posterior distribution $p(f|d=0)$ places an upper limit on the planet
occurrence frequency at a given level of confidence, $\mathrm{CL}$,
\begin{equation}\label{confidenceLimit}
\mathrm{CL} = \int_{0}^{f_{\mathrm{max}}} p(f|d=0) df.
\end{equation}

\subsection{Mapping Occurrence Frequency in the Mass-Semimajor Axis Plane}\label{mapSection}

For each of our sets of sensitivity maps \----corresponding to the three
different evolutionary models we used\---- we create maps that show our
95\% confidence upper limit to the planet frequency. That is, for each point
($a$,$m$) in our grid, with $a$ the semi-major axis and $m$ the planet mass, we
calculate a likelihood function according to Equation \ref{likelihood},
substituting $s_{i}(a,m)$ for $s_i$. Then, we use Equations \ref{posterior} and
\ref{confidenceLimit} to derive our 95\% confidence limit on the planet
occurrence frequency. For this purpose, we assume a uniform prior similar to other studies
\citep[e.g.,][]{Lafreniere2007}.

\begin{figure}[h]
\begin{center}
\includegraphics[width=0.4\textwidth]{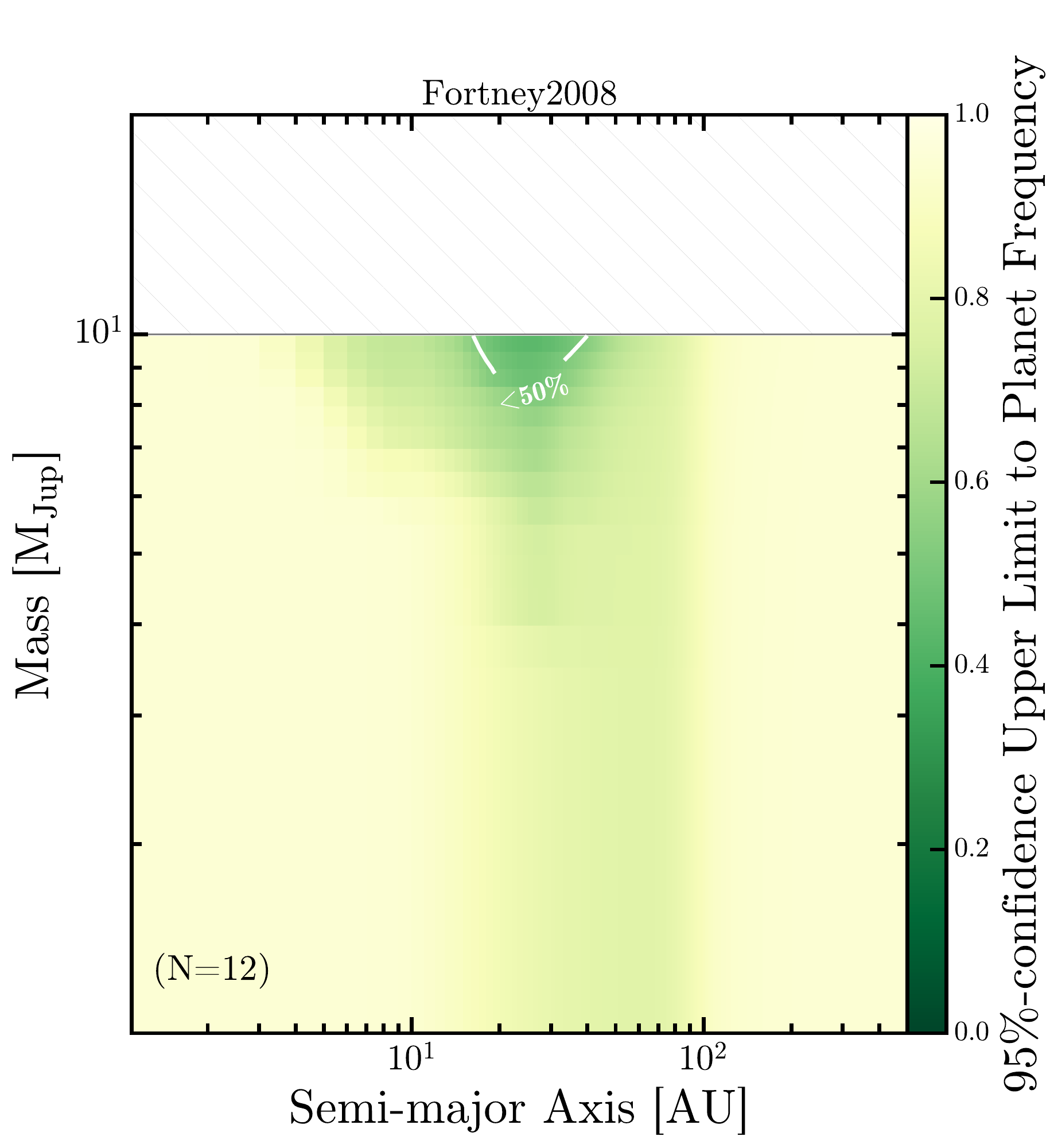} 
\caption{ LEECH 95\% confidence upper limits to the planet occurrence frequency
over a fine grid in the mass\----semimajor axis plane calculated using the F08
cold-start evolutionary models. We are sensitive to cold-start planets around
12 of our targets.  \label{FortneyLimits}} 
\end{center}
\end{figure}

The results for each evolutionary model are shown in Figures
\ref{dustySubgroups}-\ref{FortneyLimits} (DUSTY,
COND, and F08, respectively). For the DUSTY and COND constraints, in addition
to performing the analysis on the total LEECH sample, we divided the targets
into subgroups of interest. These subgroups include the subset of single stars,
the subset of FGK stars, the subset of A- and B-type stars, the subset of A and
F stars with evidence of circumstellar dust, and the subset of stars that are
members of the Ursa Major moving group. While some of these subgroups resemble
our target selection sublists, not all of them correspond directly and some
targets are members of multiple subgroups. For example, all of the dusty A stars
are included in our AB subgroup, and all FGK stars in the UMa sublist are also
included in our FGK subgroup, etc. The reordering of objects into
slightly different subgroups compared to the the target selection sublists
described in Section \ref{TargetSelection} was necessary because we did not
complete observations for all our targets, and because our dynamic nightly
scheduling resulted in uneven completion of the sublists.

Our cold-start subgroup is made up of all the stars for which we had some
sensitivity to the planets predicted by the F08 models, given our photometric
sensitivity, as well as the system age and distance. This subgroup includes
12 stars. These targets are indicated in Table \ref{TargetSummary}. 

\subsection{Comparing LEECH Maps to Previous Surveys}
\begin{figure*}
\includegraphics[width=\textwidth]{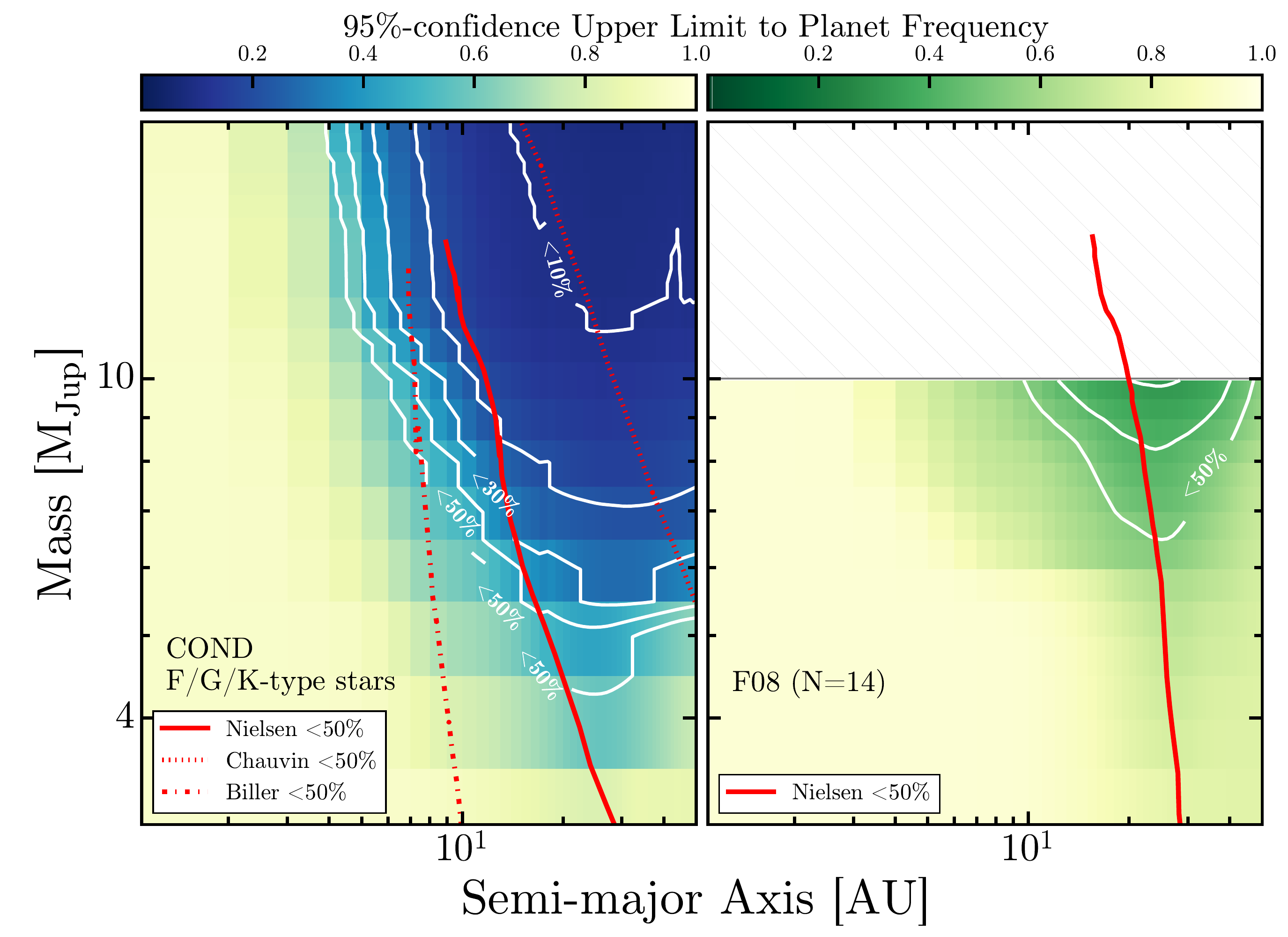}
\caption{Constraints rederived with classical $5\sigma$ contrast curves for
comparison to earlier surveys.  Left: LEECH 95\% confidence upper limits to
planet frequency around FGK stars as a function of orbital semi-major axis and
planet mass using COND evolutionary models and our classical $5\sigma$ contrast
curves, blue color map, and white contours.  We overplot, in red, the 50\%
upper-limit contour from three other surveys that publish similar maps for
target lists with similar spectral-type distributions \citep{Nielsen2010,
Biller2013, Chauvin2015}.  LEECH provides the best constraints at small
semi-major axes.  Right: Same as for the left panel but for the F08 models.
Here we compare to \citet{Nielsen2010}, who reported results for a similar
spectral-type distribution mainly consisting of FGK stars.  LEECH provides the
best constraints on cold-start planets interior to $\sim20$~au.
\label{condCompare}} 
\end{figure*}

The LEECH survey makes its most unique and significant contribution at small
separations from FGK-type stars.  To show the new parameter space probed by
LEECH, we compare our performance to previous studies. In order to make the
comparison as direct as possible, we reanalyze our data using contrast curves
created with the classical $5\sigma$ approach (not correcting for small number
statistics or varying the false-alarm rate with separation). In Figure
\ref{condCompare}, we show recalculated maps for both COND hot-start
evolutionary models and the F08 cold-start models. Our COND map is specific to
our FGK subgroup and is made using 54 stars.  We overplot the 50\% contour from
\citet{Nielsen2010}, \citet{Biller2013}, and \citet{Chauvin2015}\footnote{Our
approach used the same definition for contrast curves as used in
\citet{Nielsen2010} and \citet{Chauvin2015}. The comparison to the
\citet{Biller2013} contour is not as direct, since those authors defined their
contrast curves with a higher true-positive rate \citep[see, for example,][for
a review of the signal detection terminology]{Jensen-Clem2018}.  However, when
we compare the \citet{Biller2013} 50\% contour to the FGK map in Figure
\ref{condSubgroups}, which is based on our more rigorous contrast curves, we
still provide better constraints at small separation.}.  LEECH improves
constraints on COND-like planet occurrence frequency at small separations
($\lesssim10$--$20$~au), even though we targeted much older systems whose
hot-start planets should be intrinsically much less luminous.

For cold-start planets, only \citet{Nielsen2010} and
\citet{Brandt2014} reported limits for FGK-type stars. \citet{Nielsen2010}
published an occurrence frequency map, and we compare our LEECH results to theirs
in Figure \ref{condCompare}.  LEECH performs better interior to $\sim25$~au. We
discuss the results of \citet{Brandt2014} at more length in Section
\ref{discussionSection}.

\begin{figure}
\includegraphics[width=0.5\textwidth]{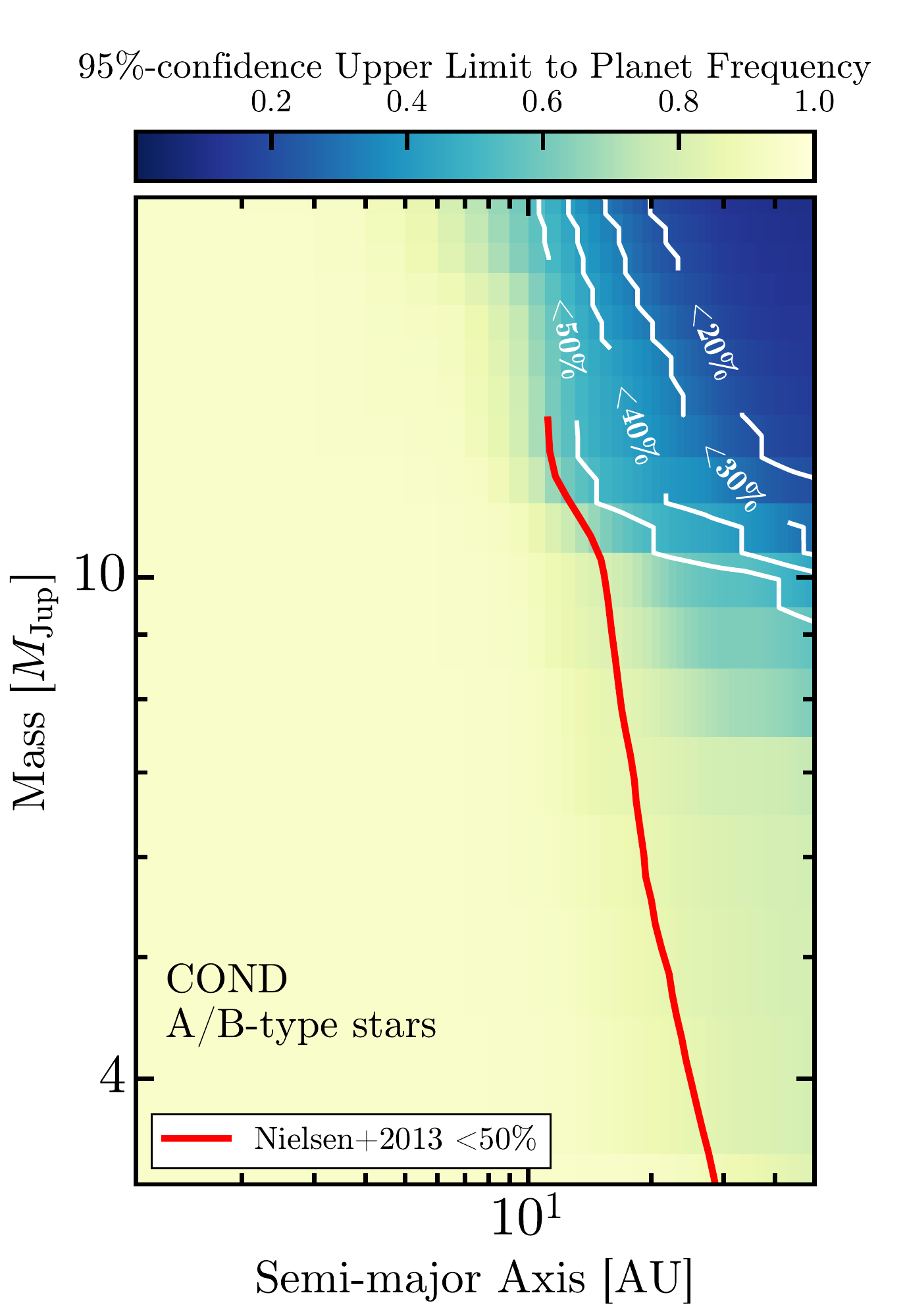}
\caption{LEECH 95\% confidence upper limits to planet frequency around A and
B stars as a function of orbital semi-major axis and planet mass using COND
evolutionary models and our classical $5\sigma$ contrast curves, blue color map,
and white contours.  We overplot, in red, the 50\% upper-limit contour from
\citet{Nielsen2013}. The LEECH 50\% contour is always within the
\citet{Nielsen2013} contour, so we do not improve upon their results. However,
LEECH typically uses older target ages than the \citet{Nielsen2013}, by
a factor of 2, reducing our derived sensitivity. \label{condABCompare}} 
\end{figure}

In Figure \ref{condABCompare} we show our map of upper limits to the planet
occurrence frequency for A- and B-type stars made with classical $5\sigma$
contrast curves. We overplot the 50\% contour of a similar map from
\citet{Nielsen2013}. In this case, the LEECH 50\% contour is always within the
\citet{Nielsen2013} contour, so we do not improve constraints on the planet
occurrence frequency compared to \citet{Nielsen2013}. However, as discussed in
Section \ref{TargetSelection} we used a different set of model isochrones than
\citet{Nielsen2013} when estimating the ages of our A- and B-type targets, and
our age estimates are typically two times as large as those found by
\citet{Nielsen2013}. This will make LEECH appear less sensitive to planets
around these types of stars.

Many of our UMa targets were also observed by \citet{AmmlervonEiff2016} as part
of the $K$-band coronagraphic imaging survey. While \citet{AmmlervonEiff2016}
considered group ages spanning 100 to 1000 Myr, they report mean sensitivity
$\lesssim5\%$ to $20~M_{\mathrm{Jup}}$ COND-like objects assuming a group age
of 500 Myr. LEECH delivers 30\% mean sensitivity to $10~M_{\mathrm{Jup}}$
objects.

We note that there are some compilation studies that perform statistical
analysis on large target lists using as input contrast curves from multiple
studies \citep[e.g.,][]{Brandt2014, Galicher2016, Bowler2016}. We cannot
compare directly to those studies in Figures \ref{condCompare} and
\ref{condABCompare} because they do not provide a similar occurrence map.
However, published compilation analyses typically provide much stronger
constraints than we show because they use a much larger number of stars. We
will discuss their results in comparison to LEECH in Section
\ref{discussionSection}. Figure \ref{condCompare} suggests LEECH contrast
curves will improve planet frequency constraints on solar system scales in
future compilation studies.

\subsection{Occurrence Frequency for Ranges of Mass and Separation} 

To place an upper limit on the occurrence frequency of planets within a range
of masses and separations, we again make use of Equations \ref{likelihood},
\ref{posterior}, and \ref{confidenceLimit}, but now we define our sensitivity
to planets for each star, $s_{i}$, to be the fraction of all Monte Carlo
injected planets over the whole range that we would detect with our LEECH data.

Figure \ref{ModelSensitivities} illustrates that we are more likely to detect
companions in certain regions of parameter space than others. Thus, a single
constraint on the planet frequency over a wide range in parameter space will
depend on our assumed shape of the underlying planet distribution. For example,
if we assume there should be lots of planets where we have a good chance of
detecting them, then we can put a better constraint on the occurrence frequency
than if we assume most planets are in regions where we are unlikely to detect
them.

For a population of planets assumed to be uniform over our grid, our
sensitivity over a range of masses and semi-major axes is simply the average
sensitivity in that range. For a non-uniform underlying planet distribution,
our sensitivity over a range of masses and separations must be calculated using
a weighted average over our sensitivity grids---essentially scaling the number
of injected planets at each grid-point as if the original Monte Carlo analysis
was performed using the assumed shape of the planet frequency distribution
\citep[e.g.,][]{Kasper2007}.  That is, we take the sensitivity to planets with
masses from $m_{\mathrm{min}}$ to $m_{\mathrm{max}}$ and semi-major axes from
$a_{\mathrm{min}}$ to $a_{\mathrm{max}}$ to be
\begin{equation} \label{sensitivityRange}
s_{i,\mathrm{range}} = \frac{\sum_{m_{\mathrm{min}}}^{m_{\mathrm{max}}}
            \sum_{a_{\mathrm{min}}}^{a_{\mathrm{max}}} w(a,m) s_{i}(a,m)}
           {\sum_{m_{\mathrm{min}}}^{m_{\mathrm{max}}}
            \sum_{a_{\mathrm{min}}}^{a_{\mathrm{max}}} w(a,m)},
\end{equation}
where $s_{i}(a,m)$ is the sensitivity map for star $i$, described in Section
\ref{sensitivitySection}, and $w(a,m)$ is the weight function that assumes the
shape of the underlying planet distribution.  Different authors make different
assumptions about the shape of the wide-orbit giant-planet distribution in
their analyses. For example, \citet{Meshkat2017} assumed a log-uniform
distribution, similar to that seen for binary stars in some cases
\citep[e.g.,][]{Duchene2013},
\begin{equation}
\frac{\partial^{2} f}{\partial\ln a\partial\ln m}\propto m^{0}a^{0}, 
\end{equation}
and \citet{Rameau2013} and \citet{Galicher2016} presented results
for a distribution that has the same power-law indices as measured for close-in
giant planets
\citep{Cumming2008}, 
\begin{equation}
\frac{\partial^{2}f}{\partial\ln a\partial\ln m}\propto m^{-0.31}a^{0.39}.
\end{equation}
and for a uniform distribution, which is not physically motivated, but rather
the default when sensitivity maps are created with uniform gridding and weights
are not used in Equation \ref{sensitivityRange},
\begin{equation}
\frac{\partial^{2}f}{\partial\ln a\partial\ln m}\propto m^{1}a^{1},
\end{equation} 
Some authors have moved beyond the power-law model, particularly for the more
well-constrained giant-planet population around M-stars.  For example,
\citet{Meyer2018} used a log-normal distribution in semi-major axis for planets
from 1 to 10 $M_{\mathrm{Jup}}$,
\begin{equation}
\frac{df}{d\ln a} \propto e^{\left(\frac{\ln a - \mu}{2\sigma^{2}}\right)}
\end{equation}
with $\mu$ and $\sigma$ the mean and variance of the log-normal, respectively.

In Table \ref{frequencyTable} we show our upper limits to planet frequency for
multiple ranges in mass and semi-major axis. For each range, we report results
for three different assumed planet distributions: (1) a power law with indices
taken to match those derived for close-in giant planets \citep{Cumming2008}, (2)
a log-uniform distribution, and (3) a uniform distribution. 

\begin{rotatetable*}
\begin{deluxetable*}{l||ccc|ccc||ccc|ccc||cc}
\tabletypesize{\footnotesize}
\tablecolumns{15}
\tablewidth{0pt}
\tablecaption{95\% Upper Limits to Planet Frequency\label{frequencyTable}}
\tablehead{
    \colhead{Subgroup}&
    \multicolumn{6}{c}{DUSTY}&
    \multicolumn{6}{c}{COND}&
    \multicolumn{2}{c}{F08}
    \\
    \colhead{}&
    \multicolumn{3}{c}{5-50 au}&
    \multicolumn{3}{c}{5-100 au}&
    \multicolumn{3}{c}{5-50 au}&
    \multicolumn{3}{c}{5-100 au}&
    \colhead{5-50 au}&
    \colhead{5-100 au}
    \\
    \colhead{}&
    \colhead{4-14$M_{\mathrm{J}}$}&
    \colhead{8-14$M_{\mathrm{J}}$}&
    \colhead{4-20$M_{\mathrm{J}}$}&
    \colhead{4-14$M_{\mathrm{J}}$}&
    \colhead{8-14$M_{\mathrm{J}}$}&
    \colhead{4-20$M_{\mathrm{J}}$}&
    \colhead{4-14$M_{\mathrm{J}}$}&
    \colhead{8-14$M_{\mathrm{J}}$}&
    \colhead{4-20$M_{\mathrm{J}}$}&
    \colhead{4-14$M_{\mathrm{J}}$}&
    \colhead{8-14$M_{\mathrm{J}}$}&
    \colhead{4-20$M_{\mathrm{J}}$}&
    \colhead{7-10$M_{\mathrm{J}}$}&
    \colhead{7-10$M_{\mathrm{J}}$}}
\startdata
\cutinhead{Cumming 2008: $\frac{\partial^{2}f}{\partial\ln a\partial\ln m}\propto m^{-0.31}a^{0.39}$}
all &19.7\% &11.1\% &15.0\% &19.7\% &10.7\% &14.6\% &24.7\% &13.9\% &18.7\% &25.3\% &13.9\% &18.5\%& 71.4\% &75.7\%\\
FGK &26.0\% &14.9\% &21.0\% &27.6\% &15.6\% &22.1\% &31.7\% &17.8\% &25.0\% &33.9\% &18.8\% &26.3\%& 87.7\% &90.3\%\\
AB  &78.7\% &52.9\% &60.3\% &67.2\% &38.4\% &46.6\% &86.1\% &72.0\% &75.7\% &79.3\% &57.2\% &62.7\%& \nodata&\nodata\\
AF  &80.1\% &67.7\% &73.3\% &72.1\% &55.1\% &62.8\% &84.3\% &76.8\% &79.7\% &78.1\% &67.3\% &70.9\%& \nodata&\nodata\\
UMa &60.0\% &31.8\% &47.0\% &60.5\% &30.5\% &45.9\% &71.2\% &39.0\% &56.6\% &72.6\% &38.5\% &56.5\%& \nodata&\nodata\\
\cutinhead{Log-uniform Distribution: $\frac{\partial^{2}f}{\partial\ln a\partial\ln m}\propto m^{0}a^{0}$}
all &20.1\% &12.2\% &15.1\% &19.6\% &11.6\% &14.3\% &25.1\% &15.2\% &18.7\% &25.0\% &14.8\% &18.0\%& 73.6\% &76.7\%\\
FGK &26.6\% &16.2\% &21.1\% &27.0\% &16.4\% &21.2\% &32.5\% &19.5\% &25.0\% &33.2\% &19.7\% &25.2\%& 88.6\% &90.1\%\\
AB  &81.7\% &60.9\% &63.2\% &72.5\% &46.3\% &50.0\% &87.9\% &78.0\% &78.3\% &83.2\% &66.3\% &67.1\%& \nodata&\nodata\\
AF  &84.3\% &75.9\% &78.3\% &78.2\% &64.7\% &68.9\% &87.5\% &82.7\% &83.6\% &83.2\% &75.4\% &76.7\%& \nodata&\nodata\\
UMa &60.5\% &35.0\% &46.6\% &59.5\% &33.1\% &44.6\% &71.5\% &42.9\% &56.0\% &71.2\% &41.5\% &54.5\%& \nodata&\nodata\\
\cutinhead{Uniform Distribution: $\frac{\partial^{2}f}{\partial\ln a\partial\ln m}\propto m^{1}a^{1}$}
all &12.9\% &9.2\%  & 9.0\% &13.4\% & 9.3\% & 8.9\% &16.3\% &11.6\% &11.2\% &17.6\% &12.3\% &11.3\%& 67.0\% &74.5\%\\
FGK &17.6\% &12.8\% &13.6\% &20.2\% &14.6\% &15.3\% &21.0\% &15.0\% &15.7\% &24.4\% &17.4\% &17.7\%& 85.3\% &90.7\%\\
AB  &54.8\% &38.3\% &30.8\% &41.0\% &27.5\% &23.2\% &72.0\% &57.4\% &45.4\% &58.4\% &42.6\% &33.4\%& \nodata&\nodata\\
AF  &63.8\% &53.1\% &49.2\% &54.4\% &42.7\% &39.9\% &72.1\% &64.6\% &59.1\% &64.1\% &55.3\% &49.1\%& \nodata&\nodata\\ 
UMa &38.4\% &26.3\% &28.1\% &39.7\% &26.2\% &27.6\% &47.3\% &32.0\% &33.3\% &50.7\% &33.1\% &33.9\%& \nodata&\nodata\\ 
\enddata
\end{deluxetable*}
\end{rotatetable*}

\section{Discussion} \label{discussionSection}
\subsection{Planet Frequency Interior to $100$~au}

\begin{figure*}[h]
\begin{center}
\includegraphics[width=\textwidth]{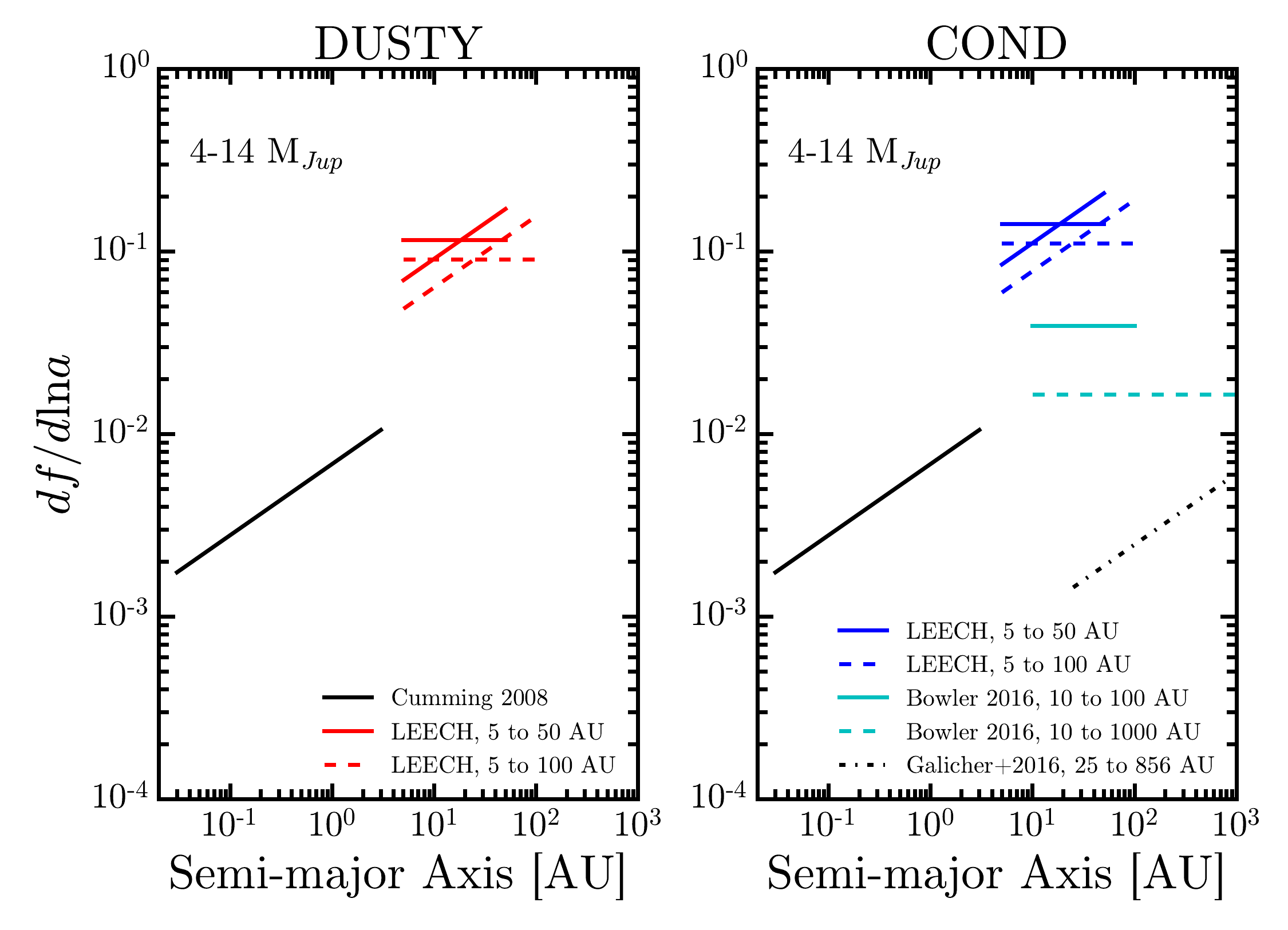}
\caption{LEECH constraints on power-law planet distributions outside 5 au
around FGK stars using hot-start planetary evolution models. All curves assume
either a log-uniform semi-major axis distribution of the planet population
(horizontal) or a power-law distribution with the same indices as found for
the RV planets interior to 3~au but allowing for a different normalization
(inclined). Upper limits are shown with colored curves and  best-fits are shown
in black. We use the LEECH constraints on planet fraction to derive an upper
limit to the normalization for an assumed power-law distribution as explained
in the text according to Equation \ref{powerlawNorm}, and then integrate over
masses from 4 to 14 $M_{\mathrm{Jup}}$ to yield a function of separation.
Left: our constraints specific to the DUSTY hot-start models are shown in red.
The solid red curves indicate our constraints using LEECH results from 5 to
50~au, and the dashed curves indicate constraints using LEECH results from 5 to
100~au. Right: LEECH upper limits specific to the COND evolutionary models are
shown in blue for the same semi-major axis ranges as the left panel. We also
plot, with cyan curves, the upper limits derived by \citet{Bowler2016} using
log-uniform distributions from 10 to 100 au (solid) and 10 to 1000 au (dashed).
The reported best fit distribution for wide-orbit planets from 4 to 14
$M_{\mathrm{Jup}}$ and from 25 to 856~au from \citet{Galicher2016} is shown
with a black dotted-dashed curve.}\label{comparison1} \end{center} 
\end{figure*}

\begin{figure}[h]
\includegraphics[width=0.5\textwidth]{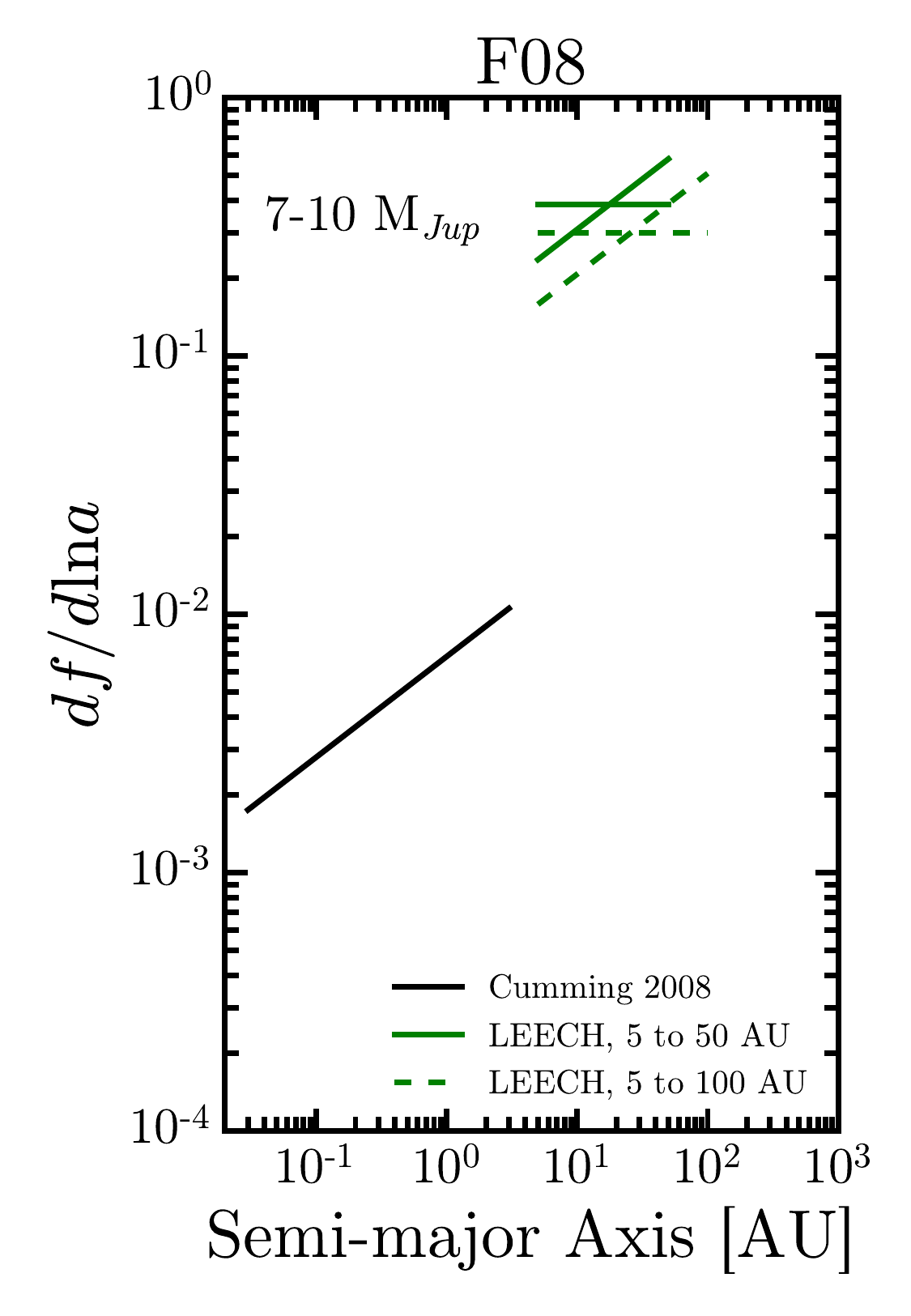}
\caption{Same as for the plots in Figure \ref{comparison1} but specific to the
cold-start models of \citet{Fortney2008}.}\label{comparison2} 
\end{figure}

It is reasonable to expect that the structure of protoplanetary disks will
affect the mass and semi-major axis distributions of giant planets. For
example, increased surface density of solids beyond the snow line could easily
result in differently shaped planet populations interior and exterior to
$\sim3$--$5$~au. Similarly, beyond the outer radii of protoplanetary disks, we
might expect very few planets. Recent, unbiased surveys with ALMA indicate that
the dust-disk radii in a typical star-forming region are $\lesssim40$~au with
very weak dependence on stellar mass \citep{Eisner2018}. Dust disks in less
common, more diffuse star forming regions are larger, and size correlates more
strongly with stellar mass in these environments; however, dust disks rarely
extend beyond  $\sim100$~au \citep[e.g.,][]{Tazzari2017}. Gas disks do extend
beyond the observed dust, but only by a factor $\sim2$ \citep{Ansdell2018}.  As
a result, constraints on planet occurrence frequency derived assuming
distributions that extend well beyond $\sim100$~au are likely underestimated.
Furthermore, recent observations of shock-tracing H$\alpha$ emission from
accreting young planets \citep{Sallum2015,Wagner2018} suggests that hot-start
evolutionary models are overly optimistic in their luminosity predictions. As
a result, constraints on planet occurrence frequency derived assuming hot-start
models are likely underestimated.  More physically meaningful constraints
should be derived using cold-start models and semi-major axis ranges better
matched to where we expect planets (e.g., interior to outer dust-disk radii).

Given a model for the giant-planet population distribution, our upper
limits in Table \ref{frequencyTable} are effectively constraints on the
normalization of the model. That is, for a power-law model,
\begin{equation}\label{powerlawNorm}
f_{\mathrm{upper}} = \int_{a_{\mathrm{min}}}^{a_{\mathrm{max}}}
                     \int_{m_{\mathrm{min}}}^{m_{\mathrm{max}}} 
                     Na^{\beta}m^{\alpha}
                     \,d\mathrm{ln}m\,d\mathrm{ln}a
\end{equation}
with $f_{\mathrm{upper}}$, $\alpha$, and $\beta$ fixed, we can solve for $N$,
the largest admissible normalization of the distribution allowed by the data.

Figures \ref{comparison1} and \ref{comparison2} graphically depict our upper
limits for FGK-type systems. Figure \ref{comparison1} shows our results
associated with hot-start models and is specific to planets from
4 to 14 $M_{\mathrm{Jup}}$.  Figure \ref{comparison2} shows our results
  associated with the cold-start \citet{Fortney2008} models and applies to the
mass range of 7 to 10 M$_{\mathrm{Jup}}$. In both figures, we show
$\frac{df}{d\mathrm{ln}a}$ as a function of orbital semi-major axis.  Each plot
includes colored curves indicating upper limits. Their slopes show the assumed
shape of the underlying planet population (\citet{Cumming2008} indices or
log-uniform), and their horizontal extent shows the corresponding semi-major
axis range used to derive the frequency upper limit.  Figures \ref{comparison1}
and \ref{comparison2} make it clear that our derived constraints on planet
frequency depend sensitively on the evolutionary model used (e.g., hot vs.
cold) and on the assumed shape of the underlying distribution, including its
radial extent.

We highlight two semi-major axis ranges in Figures \ref{comparison1} and
\ref{comparison2}: 5 to 50~au, corresponding to the domain of the Solar System
giants; and 5 to 100~au, bracketing the orbits of the imaged planets in the
HR~8799 system. In each plot and for each power-law slope, using the wider
range forces a lower normalization of the population distribution.  This is due
to the assumption that the underlying shape of the planet population remains
fixed throughout the range considered, which means that regions of
high sensitivity within the range affect the normalization throughout the
range.  We provide additional examples of this effect from the literature in
the right panel of Figure \ref{comparison1}.  \citet{Bowler2016} provided an
upper limit to 5--13 $M_{\mathrm{Jup}}$ planets over two ranges of orbital
separation, 10--100~au ($<6.8\%$) and 10--1000~au ($<5.8\%$), assuming
a log-uniform distribution of planet masses and orbital semi-major axes for the
underlying population. We use the \citet{Bowler2016} results to solve for $N$
in Equation \ref{powerlawNorm} and then adjust the upper limit for the range
4 to 14~$M_{\mathrm{Jup}}$ before plotting the corresponding curves in Figure
  \ref{comparison1}. Once again, the limiting normalization for the population
distribution is forced lower when using the wider range. The \citet{Bowler2016}
result is based on a compilation of several surveys and includes 155 FGK stars.
Compared to LEECH alone (54 FGK stars) the \citet{Bowler2016} compilation
constraint is more stringent, however \citet{Bowler2016} does not report limits
for the more narrow 5 to 50~au range, and this is where LEECH makes the
greatest contribution.

We also compare to the \citet{Galicher2016} result that 1.1\% of FGK stars have
a 4--14 $M_{\mathrm{Jup}}$ COND-like giant planet in the range from
25 to 856~au assuming a power-law distribution with indices taken from the
   close-in RV population.  The corresponding curve in Figure \ref{comparison1}
is inconsistent with an extrapolation of the close-in planet population,
implying that some change in the distribution occurs somewhere between 3 and
856 au.  Indeed, previous imaging surveys have shown that the shape of the
    short-period gas-giant distribution cannot be extended beyond $\sim60$~au
\citep[e.g.,][]{Kasper2007, Nielsen2010, Brandt2014, Reggiani2016}.  While it
is not surprising that the planet population is not a single power law from
0.03 to 856~au, it does make it difficult to interpret the 1.1\% frequency
  reported by \citet{Galicher2016}. The assumption of a single power law
from 25 to 856~au is built into the result. If instead the planet population
falls off beyond some radius, then all of the survey sensitivity outside that
radius contributes less and less to constraining the normalization of the
population (Equation \ref{sensitivityRange}). As a result, a much higher
occurrence frequency can be allowed by the data because sensitivity typically
decreases with decreasing separation for direct imaging surveys. 

We have shown that the shape of the separation distribution is crucial for
assessing the frequency of giant planets in surveys with many non-detections.
However, the shape of the gas-giant planet distribution at orbital distances of
a few times the snow line is poorly constrained. Some information is known
about the M-star planet population in this range \citep[e.g.,][]{Gould2010,
Meyer2018}, but the data suggest intrinsic differences between the M-star
planet population and the population around more massive primaries, at least at
semi-major axes that are well probed \citep[e.g.,][]{Clanton2014}.  

To minimize the effects of needing to choose an underlying planet distribution
in order to derive a planet occurrence frequency, the cautious approach is to
use the measurement sensitivity within the region of interest to infer
constraints on planet frequency for that range. In this context, the LEECH
survey has added to our ability to constrain the gas-giant planet frequency on
solar system scales by delivering sensitivity at smaller semi-major axes, a few
snow-line radii from host stars (e.g., Figure \ref{condCompare}).

\subsection{Giant Planets and Transition Disks} 
Contrary to previous results, our analysis suggests that there is no discrepancy
between the wide-gap ($\lesssim90$~au) transition disk frequency
\citep[$\gtrsim11$\%;][]{vanderMarel2018} and the frequency of wide-orbit giant
planets. For cold-start planets around solar-type stars, we know that if the
shape of the RV planet population can be extrapolated to larger orbital
separations, then it cannot extend beyond $\sim60$~au \citep{Brandt2014}.
However, if the RV planet distribution did extend to 60~au this implies
a frequency of 1 to 10 $M_{\mathrm{Jup}}$ planets on orbits from
5 to 60~au of 13.1\%, consistent with the lower limit on the transition disk
  frequency.  As we have discussed, it is unclear whether we should expect the
giant-planet distribution to continue uniformly beyond the snow line in
protoplanetary disks, so an actual constraint on the cold-start planet
population from $\sim5$ to 60~au is much more uncertain.  The LEECH constraint
is that $\lesssim88$\% of FGK stars have a 7 to 10 $M_{\mathrm{Jup}}$ planet
from 5 to 50 au, leaving open a wide range of possibilities, and we showed in
Figure \ref{condCompare} that LEECH delivers some of the best cold-start
constraints in this range.  The tension reported by \citet{vanderMarel2018}
relied on hot-start constraints for the giant-planet occurrence frequency
derived for a range of planet masses (5 to 13 $M_{\mathrm{Jup}}$) that does not
extend low enough to capture all the relevant gap-opening planets
\citep{Zhu2011, Dodson-Robinson2011}.

So, while wide-orbit planets beyond the radii of typical protoplanetary disks
are undoubtedly rare \citep[e.g.,][]{Nielsen2010}, and hot-start planets are
somewhat rare even on solar system scales \citep[e.g.,][]{Bowler2016,
Galicher2016}, it is not necessary for gas-giant planets to be rare in the
range of 5 to 100~au around solar mass primaries because they could be formed
less than maximally luminous, and in this case direct imaging surveys are far
less sensitive to detect them. Since the frequency of wide-orbit gas giants
beyond the snow line is an important parameter for studying the formation and
evolution of planetary systems, ongoing direct imaging surveys should include
a measurement of their sensitivity to cold-start planets and calculate the
corresponding occurrence frequencies or upper limits for orbital ranges
$\lesssim100$~au. 

Future observations with the James Webb Space Telescope (JWST) will
also play an important role in probing for cold-start planets on solar system
scales around nearby stars. Simulations using
pyNRC\footnote{https://pynrc.readthedocs.io} suggest that Near-Infrared Camera
(NIRCam) will be background-limited beyond $\sim3\arcsec$ when pairing the
M430R coronagraphic mask with the F430M filter, reaching $\sim19$th magnitude. 
Therefore, JWST will be capable of making background-limited probes interior to
100~au around stars within 30~pc. There are 70 FGK stars within 10 pc
\citep{Henry2018}, scaling by volume there are nearly 2000 FGK stars within 30
pc.

The NIRCam F430M filter is similar to $M^{\prime}$, allowing us to make
a direct comparison of the NIRCam background limit to the predictions for
gas-giant planet brightness from evolutionary models.  According to the COND
evolutionary models, a 19th magnitude sensitivity limit should facilitate the
detection of a $3~M_{\mathrm{Jup}}$ planet at 10 pc, or a $6~M_{\mathrm{Jup}}$
planet at 30 pc in a 5 Gyr system. The F444W filter will be employed
more often than F430M for planet searches due to its broad wavelength coverage,
providing an additional $\sim1$ mag increase in overall sensitivity
compared to F430M.

\section{Conclusion} 
We presented the results of the LEECH direct imaging survey for wide-orbit
gas-giant planets. LEECH was performed at $3.8~\mu\mathrm{m}$ where colder
planets emit more of their flux. This allowed us to emphasize proximity over
youth in our target selection, resulting in increased sensitivity interior to
20~au compared to previous surveys.

We reached deeper average contrast around our targets than \citet{Rameau2013},
who also reported the results of a large $L^{\prime}$ survey. We are typically
$\gtrsim1$ mag more sensitive as a function of angular separation due to
the performance of the LBT deformable secondary AO system and the
thermal-infrared sensitivity of LBTI/LMIRcam. We are even more sensitive as
a function of orbital radius in astronomical units after accounting for the different average
distance of our targets, 25 pc for LEECH and 40 pc for \citet{Rameau2013}.

We converted our photometric limits to limits on the minimum detectable planet
mass around each star by using evolutionary models to convert luminosity and
age to mass. We used three different evolutionary models that bracket
observations and span extreme assumptions regarding the zero-age luminosity of
planets and their atmospheric appearance.

Ages for each target were mostly taken from the literature. For our A- and
B- type field stars, we derived ages following the approach of
\citet{Nielsen2013}. Our results are systematically older than
those of \citet{Nielsen2013} for the stars in common, most likely because we use
different model isochrones.

Our survey delivers the best-yet sensitivity to cold-start planets interior to
20~au around FGK stars. We used our survey results to place constraints on the
wide-orbit giant-planet occurrence rate around these stars. We discussed how
such limits depend sensitively on the choice of evolutionary model as well as
the underlying planet distribution. We showed that when conservative choices
are made (using cold-start evolutionary models and considering only a narrow
range of semi-major axes not extending beyond the typical protoplanetary disk
radius), the giant-planet occurrence frequency on 5 to 100~au orbits is not
well constrained (Table \ref{frequencyTable}).  Planets in this range may be
common.

\acknowledgements{J.M.S. is supported by NASA through Hubble Fellowship grant
HST-HF2-51398.001-A awarded by the Space Telescope Science Institute, which is
operated by the Association of Universities for Research in Astronomy, Inc.,
for NASA, under contract NAS5-26555. LEECH is funded by the NASA Origins of
Solar Systems Program, grant NNX13AJ17G. K.M.M.'s work is supported by the NASA
Exoplanets Research Program (XRP) by cooperative agreement NNX16AD44G The
results reported herein benefited from collaborations and/or information
exchange within NASA’s Nexus for Exoplanet System Science (NExSS) research
coordination network at Arizona State University sponsored by NASA’s Science
Mission Directorate (grant NNX15AD53G). This material is based in part on work
supported by the National Aeronautics and Space Administration under Agreement
No.  NNX15AD94G, Earths in Other Solar Systems, issued through the Science
Mission Directorate interdivisional initiative Nexus for Exoplanet System
Science. This research has made use of the Washington Double Star Catalog
maintained at the U.S. Naval Observatory.  This publication makes use of data
products from the Two Micron All Sky Survey, which is a joint project of the
University of Massachusetts and the Infrared Processing and Analysis
Center/California Institute of Technology, funded by the National Aeronautics
and Space Administration and the National Science Foundation. This work has
made use of data from the European Space Agency (ESA) mission {\it Gaia}
(\url{https://www.cosmos.esa.int/gaia}), processed by the {\it Gaia} Data
Processing and Analysis Consortium (DPAC,
\url{https://www.cosmos.esa.int/web/gaia/dpac/consortium}). Funding for the
DPAC has been provided by national institutions, in particular the institutions
participating in the {\it Gaia} Multilateral Agreement.  This work benefited
from the Exoplanet Summer Program in the Other Worlds Laboratory (OWL) at the
University of California, Santa Cruz, a program funded by the Heising-Simons
Foundation.  The LBT is an international collaboration among institutions in
the United States, Italy and Germany.  LBT Corporation partners ar: The
University of Arizona on behalf of the Arizona university system; Istituto
Nazionale di Astrofisica, Italy; LBT Beteiligungsgesellschaft, Germany,
representing the Max-Planck Society, the Astrophysical Institute Potsdam, and
Heidelberg University; The Ohio State University, and The Research Corporation,
on behalf of The University of Notre Dame, University of Minnesota, and
University of Virginia. }

\appendix
\section{Constructing More Modern Contrast Curves}\label{Appendix}

We created contrast curves that account for small number statistics
\citep[e.g.,][]{Mawet2014} and provide a high level of completeness
\citep[e.g.,][]{Wahhaj2013}. Our specific approach closely followed the example
of \citet{Ruane2017}, using a varying detection threshold with separation from
the target star in order to provide a constant number of expected false
detections at each radius \citep{Jensen-Clem2018}. Our curves are designed to
deliver 95\% completeness to objects above our threshold chosen to provide 0.01
total false detections per data set. In this appendix we discuss how we
adjusted our classical $5\sigma$ contrast curves to meet these criteria.

For each target and for each radius, $r$, in units of $\lambda/D$, we
define the acceptable number of false detections, $\langle N_{
\mathrm{false} } (r) \rangle$, by dividing 0.01 total false detections per
target evenly among the 28 annuli of $\lambda/D$ width that exist in
our $3\arcsec$ field of view. We then solved for the corresponding acceptable
false-positive fraction as a function of radius, $\mathrm{FPF}(r)$, using
\begin{equation}\label{nexpected} 
\mathrm{FPF}(r) = \frac{\langle N_{\mathrm{false}}(r) \rangle}{2 \pi r}, 
\end{equation} 
where $2\pi r$ is the number of independent samples of the noise distribution
at each radius.  Thus, a larger false-positive fraction is used for smaller
separations in our curves. 

To connect our measured photometry $S$ to false-positive fractions given an
estimate of the noise level $\sigma$ (derived using $n=2\pi r$ independent
samples of the noise), we use the value 
\begin{equation} 
t = \frac{S}{\sigma\sqrt{1+1/n}}, 
\end{equation} 
which is $t$-distributed with $n-1$ degrees
of freedom, assuming the underlying noise distribution is intrinsically
Gaussian \citep[following, e.g.,][]{Ruane2017}. We then calculated the necessary
threshold for this value, $\tau(r)$,  to attain our required $\mathrm{FPF}(r)$.
To do this we used the percentage point function (or inverse cumulative
distribution, $ppf_t$), of the $t$-distribution with $n-1$ degrees of freedom,
\begin{equation}\label{ppf}
\tau(r) = ppf_{t}(1-\mathrm{FPF}(r), n-1)+ppf_{t}(0.95, n-1)
\end{equation}
where the last term on the right is necessary to ensure 95\% completeness.

\begin{figure}
\plotone{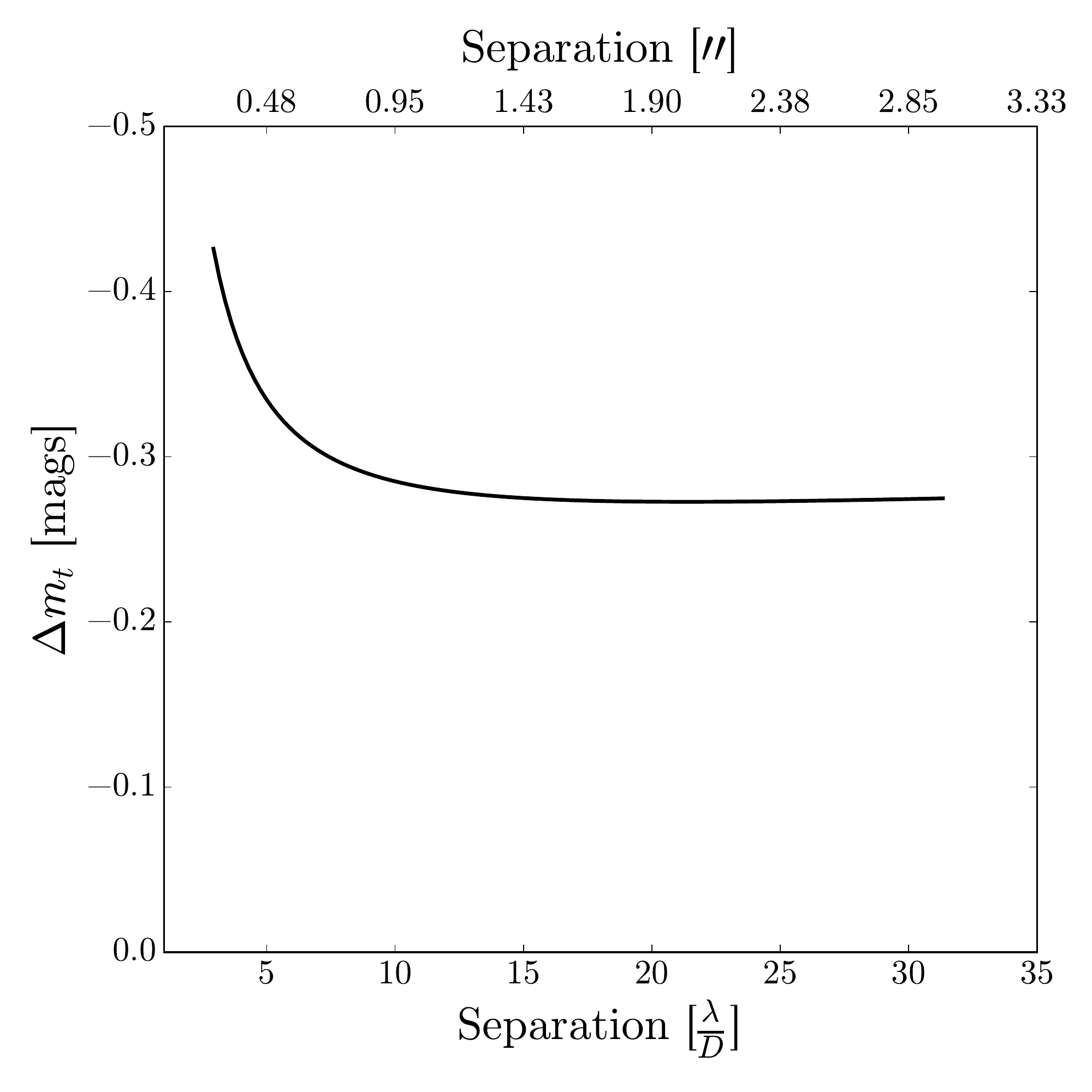} 
\caption{Contrast adjustment as a function of separation to convert our
classical $5\sigma$ contrast curves to curves that indicate a constant number
of expected false detections per radius, properly accounting for small number
statistics, and ensuring 95\% completeness.\label{offsetFig} }
\end{figure}

Finally, we adjust our contrast curves to meet this threshold by adding
\begin{equation}
\Delta m_{t}(r) = 2.5\log_{10}\left( \frac{t_{\mathrm{old}(r)}}{\tau(r)} \right),
\end{equation}
where $t_{\mathrm{old}}(r)$ is the $t$-value as a function of separation
corresponding to our classical $5\sigma$ contrast curves,
\begin{equation}
t_{old}(r) = \frac{5}{\sqrt{1+1/n}}.
\end{equation}
In Figure \ref{offsetFig}, we plot $\Delta m_{t}(r)$. The average adjustment to
our classical contrast curves is 0.29 mag.

\end{document}